\documentclass{elsarticle}
\pdfoutput=1

\usepackage{feynmf}
\usepackage{amsmath}
\usepackage{amssymb}
\usepackage{hyperref}

\newcommand{\noun}[1]{\textsc{#1}}

\makeatletter
\@ifundefined{showcaptionsetup}{}{%
 \PassOptionsToPackage{caption=false}{subfig}}
\usepackage{subfig}
\makeatother

\begin{document}
\title{Two-particle irreducible effective actions versus resummation: analytic
properties and self-consistency\tnoteref{t1}}
\author[jcu]{Michael~Brown\corref{cor1}}
\ead{michael.brown6@my.jcu.edu.au}

\author[jcu]{Ian~Whittingham}
\ead{ian.whittingham@jcu.edu.au}

\cortext[cor1]{Corresponding author}

\address[jcu]{College of Science, Technology and Engineering, James Cook University,\\
Townsville, QLD 4811, Australia}

\tnotetext[t1]{This article is registered under preprint number: /1503.08664}

\begin{abstract}
Approximations based on two-particle irreducible (2PI) effective actions
(also known as $\Phi$-derivable, Cornwall-Jackiw-Tomboulis or Luttinger-Ward
functionals depending on context) have been widely used in condensed
matter and non-equilibrium quantum/statistical field theory because
this formalism gives a robust, self-consistent, non-perturbative and
systematically improvable approach which avoids problems with secular
time evolution. The strengths of 2PI approximations are often described
in terms of a selective resummation of Feynman diagrams to infinite
order. However, the Feynman diagram series is asymptotic and summation
is at best a dangerous procedure. Here we show that, at least in the
context of a toy model where exact results are available, the true
strength of 2PI approximations derives from their self-consistency
rather than any resummation. This self-consistency allows truncated
2PI approximations to capture the branch points of physical amplitudes
where adjustments of coupling constants can trigger an instability
of the vacuum. This, in effect, turns Dyson's argument for the failure
of perturbation theory on its head. As a result we find that 2PI approximations
perform better than Pad\'e approximation and are competitive with Borel-Pad\'e
resummation. Finally, we introduce a hybrid 2PI-Pad\'e method.
\end{abstract}

\begin{keyword}
two particle irreducible effective action \sep resummation \sep nonperturbative \sep quantum field theory \sep Borel summation \sep Pad\'e approximant

arXiv: 1503.08664
\MSC[2010] 81T10, 81Q20, 81Q15, 65B10
\end{keyword}

\maketitle

\tableofcontents{}

\section{Introduction\label{sec:Introduction}}

Two-particle irreducible (2PI) effective actions and approximation
schemes based on them are often touted as useful techniques when it
is necessary to go beyond the standard perturbative field theory, with applications to thermal and non-equilibrium plasmas/fluids, strongly coupled quantum field theories and systems dominated by many-body collective effects.
The technique was originally developed by Lee and Yang \citep{Lee1960}, Luttinger and Ward \citep{Luttinger1960}, Baym \citep{Baym1962} and others in the context of many-body theory, then extended by Cornwall, Jackiw and Tomboulis \citep{Cornwall1974} to relativistic field theory where it found its natural formulation in terms of functional integrals. Since then a broad literature has developed surrounding 2PI effective actions and their generalizations (see \citep{Berges2004} for a good introductory review).

Approximations based on 2PI effective actions are often justified
as a selective re-summation of perturbation theory (some recent examples:
\citep{Berges2006,Arai2013a,Apo}; interestingly though, this point
of view is not found in the Cornwall, Jackiw and Tomboulis paper \citep{Cornwall1974}),
however they are not really a summation method in the same sense as,
for example, Borel summation. Rather, this interpretation comes in
a rather roundabout way. First, a set of self-consistent equations
of motion is derived for the mean field $\varphi=\left\langle \phi\right\rangle $
and connected correlation function $\Delta=\left\langle \phi^{2}\right\rangle -\left\langle \phi\right\rangle ^{2}$,
starting directly from the non-perturbative definition of the theory
through the path integral. Only after formally solving the equations
of motion by repeatedly iterating them does one obtain the usual perturbative
expansion for these quantities or, if the equations of motion are
truncated, a selective re-summation of perturbation theory appears.

In this work we attempt to clarify the connection between 2PI approximations,
traditional perturbation theory and re-summation methods with a special
emphasis on what analytic features of the 2PI formalism allow improved
approximations to be obtained in the presence of a divergent Feynman
diagram series. Unfortunately robust comparisons are difficult because
the large order behaviour of perturbation theory is known, at best,
in a sketchy form for most field theories of interest. To that end
we restrict attention to a genuinely trivial model ``field theory''
in zero spacetime dimensions (i.e. probability theory) for which exact
results are easily obtainable and all complications due to renormalization
etc. disappear. This model nevertheless accurately represents the
typical combinatoric structure of large order perturbation theory,
at least in those cases where the behaviour is known in more interesting
field theories. We also introduce a ``spectral function'' representation
of the Green function (similar to the one first introduced by Bender and Wu \citep{Bender1971,Bender1978})
to capture the non-analyticity of the solutions in the various methods.

The existence of the spectral representation is connected to the branch cut
of physical amplitudes on the negative coupling ($\lambda$) axis.
This branch cut is due to the non-existence of the theory at negative
couplings: the path integral diverges due to a potential unbounded from below.
In a higher dimensional field theory this has a simple physical interpretation:
the vacuum is unstable and, after tunneling through a barrier, the system rolls
down the potential \citep{Coleman1977}.
For weak coupling the semi-classical approximation is valid and the tunneling is
exponentially suppressed, giving an imaginary contribution $\sim\exp\left(-1/\lambda\right)$
to the vacuum persistence amplitude which is inherited by the Green function.
This exponential behaviour can be seen in the spectral function we obtain.

Dyson \citep{Dyson1952} argued that a very similar phenomenon occurs in quantum
electrodynamics (QED). We briefly reiterate this argument.
In QED physical observables are calculated in a perturbation series of the form
$F\left(e^2\right) = a_0 + a_2 e^2 + a_4 e^4 + \cdots$ where $e\sim 0.3$ is the charge of the electron.
Now if one imagines a world where $e^2 < 0$, i.e. like charges attract, it is easy
to see that the ordinary vacuum is unstable to the production of many
electron-positron pairs which separate into clouds of like-charged particles.
At weak coupling there is a large tunneling barrier to overcome because one
must pay for the rest mass of the pairs and separate them far enough for the
wrong-sign Coulomb potential to compensate. Thus there is a finite but
exponentially suppressed rate of vacuum decay.
A Taylor series expansion in $e^2$ cannot capture this non-analyticity so
the perturbation series must be divergent.

Similarly, the perturbation series in $\lambda$ diverges for the toy model considered here.
Pad\'e approximants are more effective because they can develop isolated poles
in the complex $\lambda$ plane, however they struggle to capture the strong
coupling behaviour at any fixed order in the approximation.
Pad\'e approximants are better able to capture the non-analyticities of the Borel
transform, however, and the widely used combination Borel-Pad\'e approximants
give a better global approximation. This occurs because the Pad\'e approximated
Bore transform has poles in the Borel plane, which lead to branch cuts when
the Laplace transform is taken to return to physical variables.
Similarly, the self-consistent 2PI approximations develop branch point
non-analyticities and approximate the exact Green function rather well in the entire
complex $\lambda$ plane already at the leading non-trivial truncation.
However, the branch cuts in the 2PI case arise because the 2PI Green function
obeys self-consistent equations of motion, and is connected to the existence
of unphysical solution branches.

A question that naturally arises is: how do these methods compare?
Both 2PI and Borel-Pad\'e methods have the ability to accurately represent
non-analyticities of the exact theory and so out-perform other methods.
However, we conjecture that self-consistently derived equations of motion
``know more'' about the analytic structure of the underlying theory than
do the generic Borel-Pad\'e approximants. Hence we test the hypothesis that
the 2PI methods should be more accurate that Borel-Pad\'e and, indeed,
find this to be the case, at least in certain regimes.

The theory discussed here, although admittedly a toy model, also has physical relevance.
Independent from us, Beneke and Moch found this toy model as the theory governing
the zero mode of scalar fields in Euclidean de Sitter space \citep{Beneke2013}.
They performed an analysis very similar to ours, finding that a non-perturbative
treatment is necessary and comparing 2PI and (Borel-)Pad\'e resummed approximations.
However, they present this analysis very briefly as part of a larger discussion
of scalar fields in de Sitter space. Further, their comparison of the 2PI and
resummed techniques, while correct as far as we can tell, is not very detailed.
Here we present detailed discussion of the interplay between 2PI effective actions
and various resummation techniques. Our use of the spectral function to quantify
the non-analyticities present in the Green function in aid of this comparison is,
as far as we are aware, a new aspect.

This paper has a somewhat pedagogical flavour, and readers familiar
with field theory, Borel summation and Pad\'e approximants can skim
through sections \ref{sec:Exact-solution-of} and \ref{sec:Perturbation-theory-and-resummations}
where these topics are discussed, pausing only to pick up our notation
and our derivation of the spectral function for the exact theory (Section
\ref{sec:Exact-solution-of}) and the Pad\'e resummed theory (Section
\ref{sub:Pade-approximation}). In Section \ref{sec:2PI-Approximations}
we compute the 2PI effective action, Green function and corresponding
spectral function for the theory. In Section \ref{sec:Hybrid-2PI-Pade}
we introduce, for the first time to our knowledge, a hybrid 2PI-Pad\'e
scheme and show that it accelerates convergence at weak coupling while
having the correct behaviour at strong coupling, like 2PI approximations
but not the usual Pad\'e method. Finally in Section \ref{sec:Discussion}
we draw our conclusions.

\section{Exact solution of zero dimensional QFT\label{sec:Exact-solution-of}}

We consider a Euclidean QFT in zero dimensions, i.e. a probability
theory for a single real variable $q$ given by the partition function
\begin{equation}
Z\left[K\right]=N\int_{-\infty}^{\infty}\mathrm{d}q\exp\left(-\frac{1}{2}m^{2}q^{2}-\frac{1}{4!}\lambda q^{4}-\frac{1}{2}Kq^{2}\right)\label{eq:partition-function}
\end{equation}
in the presence of a source $K$ for the two point function. $N$
is a normalizing factor chosen so that $Z\left[0\right]=1$. This
theory has been discussed before in the context of exact and non-perturbative
methods in field theory (see, e.g., \citep{Malbouisson1999,Crutchfield1979}
and references therein) because, despite the absence of spacetime,
the theory possesses a perturbative expansion in terms of Feynman
diagrams with the same combinatorial structure as more realistic theories.
It was also discussed in \citep{Beneke2013} as an effective field theory
for scalar field zero modes in Euclidean de Sitter space.

We restrict attention to the $m^{2}>0$ theory since, though the $m^{2}<0$ theory
exists and may be interesting for other purposes, it possesses no
sensible weak coupling limit and in zero dimensions does not give
a broken symmetry phase anyway.\footnote{$Z\left[K\right]$ only really depends on the ratio $\lambda/m^{4}$,
so the only sensible definition of weak coupling is $\lambda\ll m^{4}$,
a limit which sends the entire support of the integral to $q\to\pm\infty$
if $m^{2}<0$.}\textsuperscript{,}\footnote{The standard argument for spontaneous symmetry breaking, i.e. that
the tunneling amplitude between vacua tends to zero exponentially as the
volume of spacetime tends to infinity, is clearly inapplicable in this
case.} In the absence of symmetry breaking we may omit a source term for
$q$. The integral diverges for $\mathrm{Re}\lambda<0$, but we can
define the value of $Z\left[K\right]$ at complex $\lambda$ by analytic
continuation. Then $Z\left[K\right]$ possesses a branch cut along
the negative $\lambda$ axis. Physically, this signals the instability
of the negative $\lambda$ vacuum due to tunneling away from the
local minimum at $q\sim0$ to $q\sim\pm\sqrt{-6m^{2}/\lambda}$ followed
by rolling down the inverted quartic potential which is unbounded
from below. The branch point at $\lambda=0$ means that the weak coupling
perturbation series has zero radius of convergence. This is similar
to the behaviour which exists in most theories of physical interest
as argued by Dyson \citep{Dyson1952}.

\begin{figure}
\subfloat[\label{fig:z-integrand-abs}]{\protect\includegraphics[height=0.4\textheight]{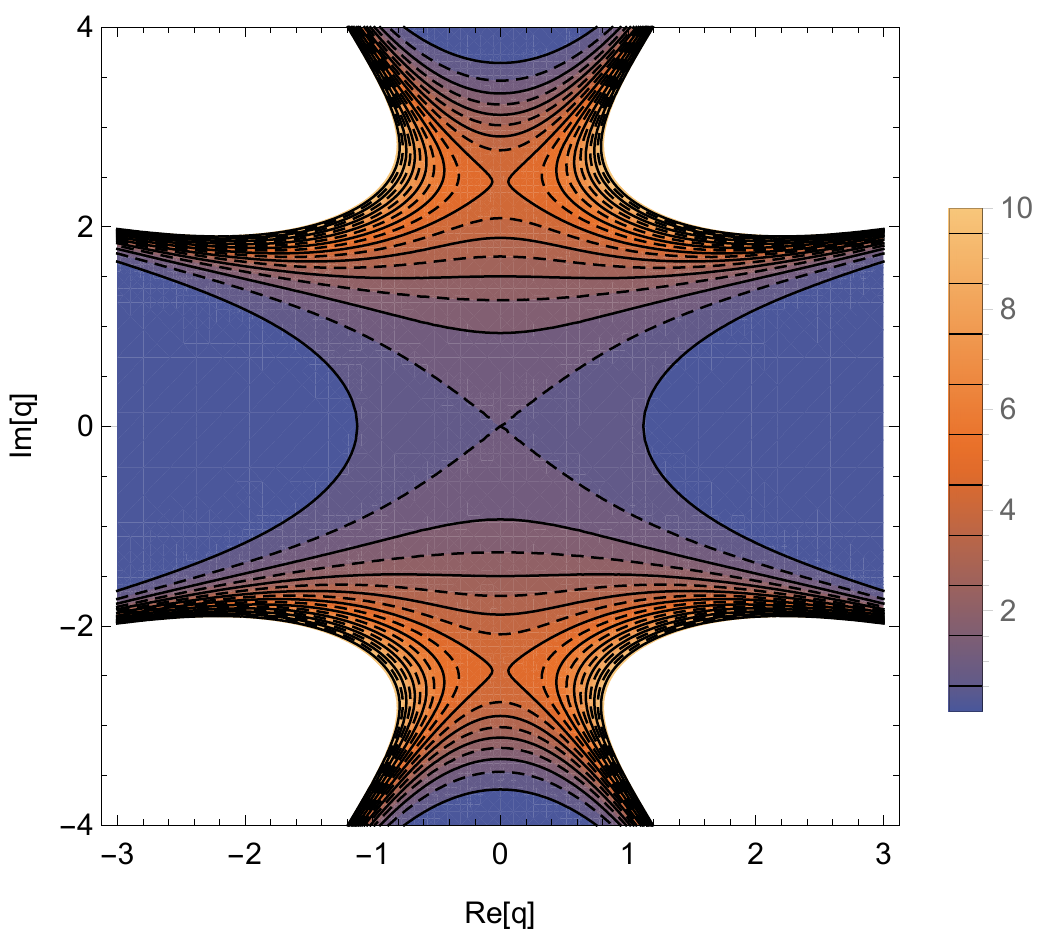}

}

\subfloat[\label{fig:z-integrand-arg}]{\protect\includegraphics[height=0.4\textheight]{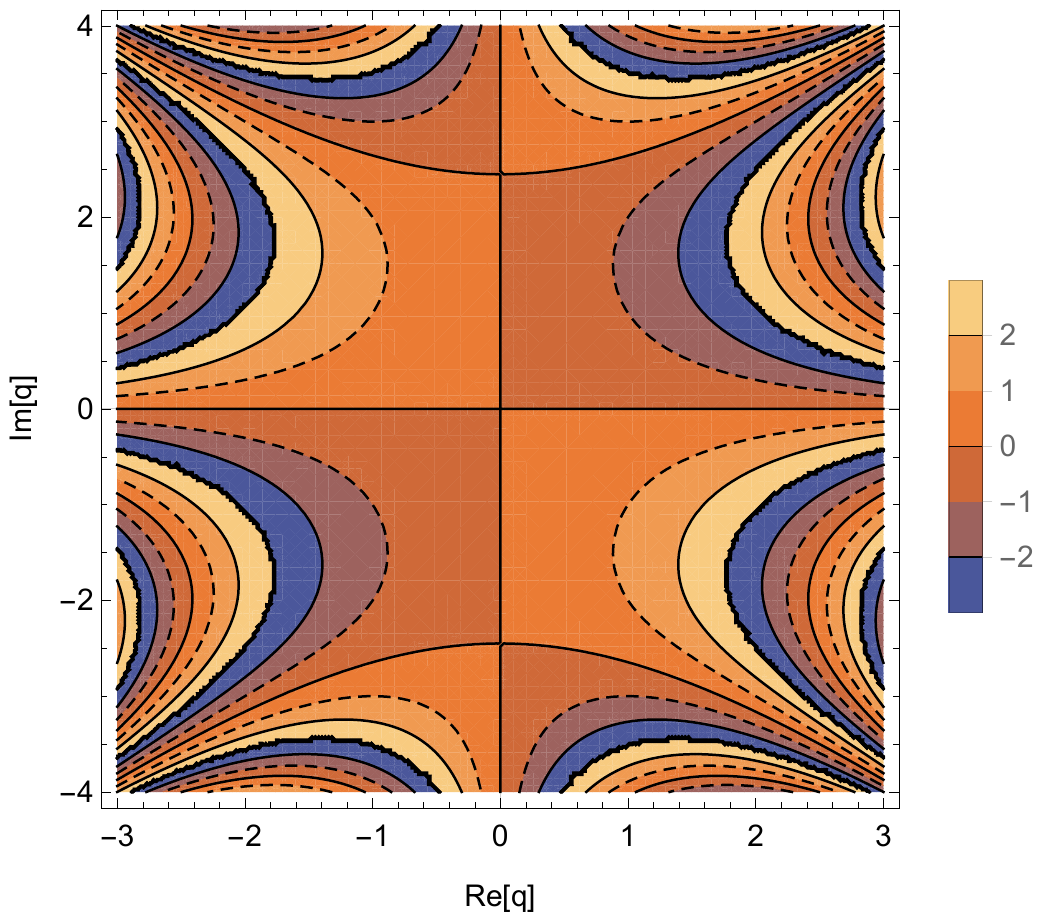}

}

\protect\caption{\label{fig:z-integrand}Modulus \ref{fig:z-integrand-abs} and phase
\ref{fig:z-integrand-arg} of the integrand in \eqref{eq:partition-function}
for $m^{2}+K=1$, $\lambda=1$. The integrand has a maximum at $q=0$
and saddles at $q=\pm im\sqrt{6/\lambda}$.}

\end{figure}

The analytic behaviour of the integrand for $m^{2},\lambda>0$ is
shown in Figure \ref{fig:z-integrand}. One can see that the integrand
has a maximum at $q=0$ and saddle points at $q=\pm im\sqrt{6/\lambda}$.
$Z\left[K\right]$ for the $m^{2}>0$ theory can be obtained from integration
along the real axis, while integration along the imaginary axis gives
the $m^{2}<0$ theory. Similarly, taking the reciprocal of the integrand
sends $m^{2}\to-m^{2}$, $\lambda\to-\lambda$ and also reverses the
colour map in Figure \ref{fig:z-integrand} (changing maxima to minima
and vice versa). From this we can see that, as expected, the partition
function diverges for $\lambda<0$ and theory makes no sense.

We introduce the conveniently rescaled variables $k=K/m^{2}$ and
$\rho=3m^{4}/4\lambda$ and obtain
\begin{equation}
Z\left[K\right]=\frac{2N}{m}\sqrt{\rho\left(1+k\right)}\exp\left(\rho\left(1+k\right)^{2}\right)\mathrm{K}_{1/4}\left(\rho\left(1+k\right)^{2}\right),
\end{equation}
where $K_{1/4}\left(\cdots\right)$ is a modified Bessel function
of the second kind. This expression is valid so long as
\begin{equation}
\mathrm{Re}\left(\sqrt{\rho}\left(1+k\right)\right)>0,
\end{equation}
which extends the definition \eqref{eq:partition-function} to the
entirety of the cut $\lambda$-plane.
The normalization factor is
\begin{equation}
N=\frac{m}{2}\frac{\exp\left(-\rho\right)}{\sqrt{\rho}\mathrm{K}_{1/4}\left(\rho\right)},
\end{equation}
and finally
\begin{equation}
Z\left[K\right]=\sqrt{1+k}\exp\left(\rho\left[\left(1+k\right)^{2}-1\right]\right)\frac{\mathrm{K}_{1/4}\left(\rho\left(1+k\right)^{2}\right)}{\mathrm{K}_{1/4}\left(\rho\right)}.
\end{equation}

$Z\left[K\right]$ is used to compute the expectation value of physical
observables $O\left(q^{2}\right)$, by the standard trick of differentiating
under the integral and then removing the source:
\begin{align}
\left\langle O\left(q^{2}\right)\right\rangle  & \equiv N\int_{-\infty}^{\infty}\mathrm{d}q\ O\left(q^{2}\right)\exp\left(-\frac{1}{2}m^{2}q^{2}-\frac{1}{4!}\lambda q^{4}\right)\nonumber \\
 & =\left.O\left(-2\partial_{K}\right)Z\left[K\right]\right|_{K=0}.
\end{align}
We also define the generating function
\begin{equation}
W\left[K\right]=-\ln Z\left[K\right]\label{eq:w-generating-function}
\end{equation}
and note that averages are found by taking derivatives of $W$. For
example,
\begin{align}
\frac{\partial}{\partial K}W\left[K\right] & =\frac{1}{2}\left\langle q^{2}\right\rangle _{K}\equiv\frac{1}{2}\bar{G},\label{eq:g-def}\\
\frac{\partial^{2}}{\partial K^{2}}W\left[K\right] & =-\frac{1}{4}\left(\left\langle q^{4}\right\rangle -\left\langle q^{2}\right\rangle ^{2}\right)_{K}\equiv\frac{1}{4}\left(V^{\left(4\right)}\bar{G}^{4}-2\bar{G}^{2}\right),\label{eq:4pt-vertex-def}
\end{align}
where the subscript $K$ indicates the average is taken at a fixed
value of $K$. (Note that $W\left[K\right]$ is not the \emph{connected}
generating function as usually defined because $K$ is a \emph{two}-point
source.) $\bar{G}$ and $V^{\left(4\right)}$ are the proper two and
four point functions respectively. To lowest order in perturbation
theory and with $K=0$, $\bar{G}=1/m^{2}+\mathcal{O}\left(\lambda\right)$
and $V^{\left(4\right)}=\lambda+\mathcal{O}\left(\lambda^{2}\right)$.

\begin{figure}
\includegraphics[width=1\columnwidth]{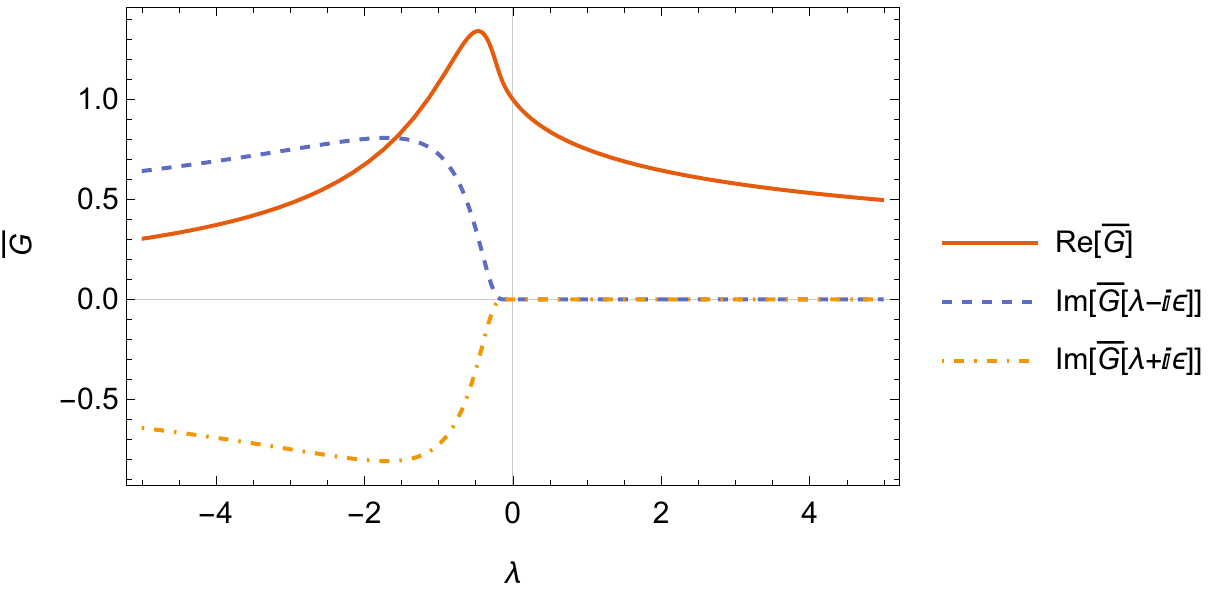}

\protect\caption{\label{fig:Gexact}$\bar{G}$ from \eqref{eq:exact-G-soln} as a function
of $\lambda$ for $m=1$. The sign of the imaginary part depends on
whether one approaches the $\lambda<0$ cut from above or below. Note the imaginary part
is exponentially suppressed near the origin because the vacuum decay process is non-perturbative.}
\end{figure}

\begin{figure}
\subfloat[\label{fig:Gexact-abs}]{\protect\includegraphics[height=0.4\textheight]{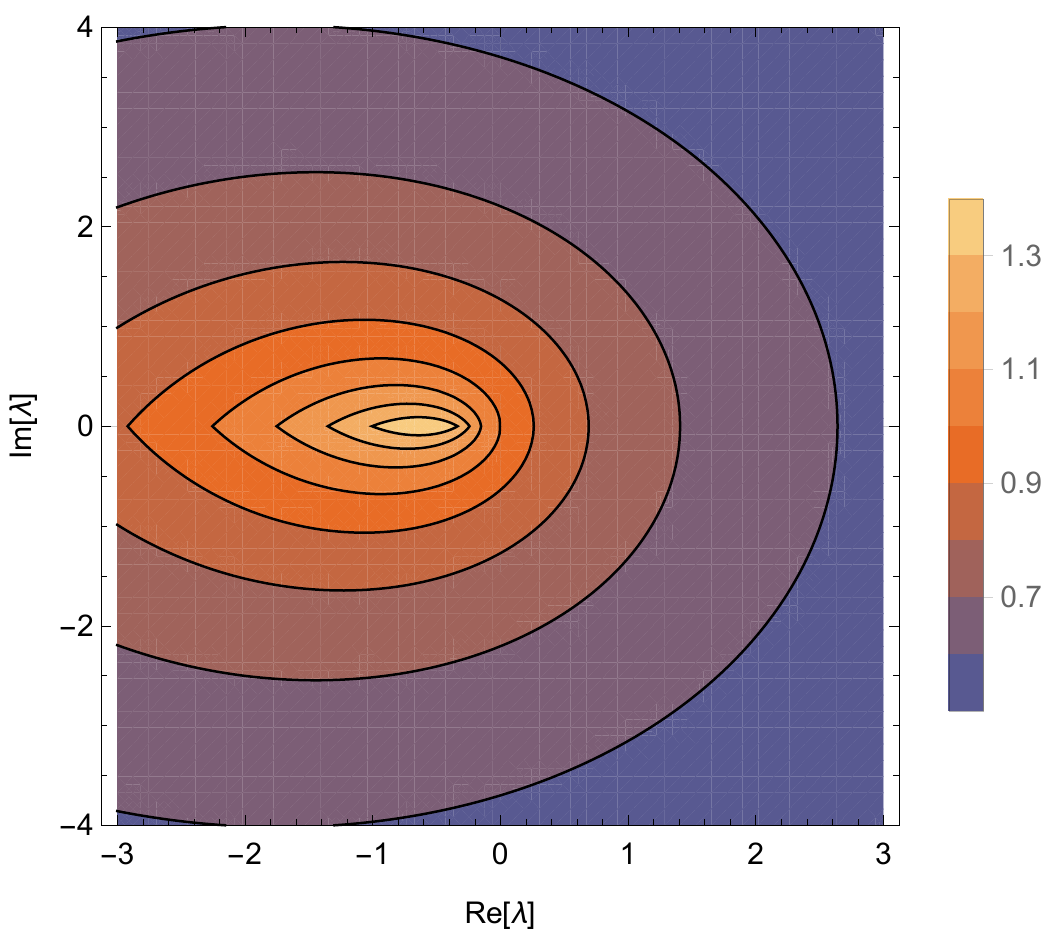}

}

\subfloat[\label{fig:Gexact-arg}]{\protect\includegraphics[height=0.4\textheight]{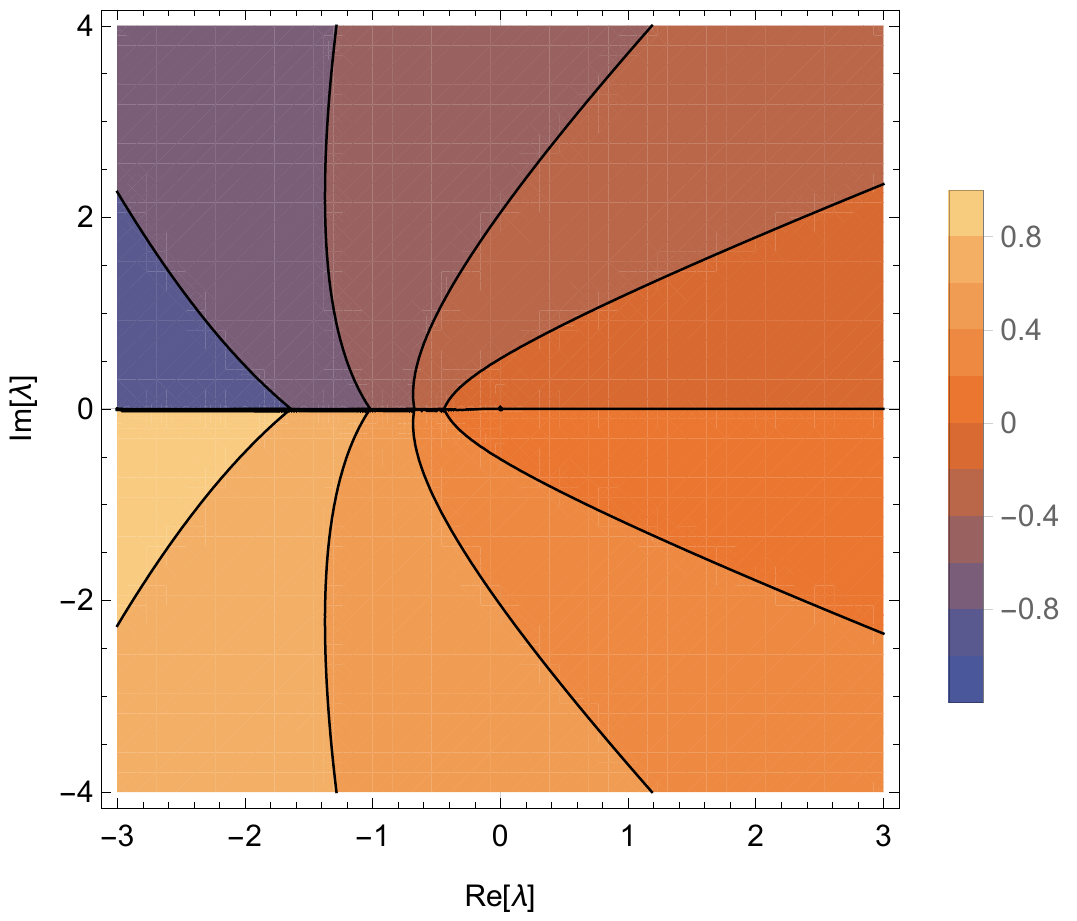}

}

\protect\caption{\label{fig:Gexact-analytic-behaviour}Modulus \ref{fig:Gexact-abs}
and phase \ref{fig:Gexact-arg} of $\bar{G}$ in \eqref{eq:exact-G-soln}
for $m^{2}=1$ in the complex $\lambda$ plane.}
\end{figure}

The exact value of $W\left[K\right]$ is easily obtained directly
from the definition, giving
\begin{align}
W\left[K\right] & =-\frac{1}{2}\ln\left(1+k\right)-\rho\left[\left(1+k\right)^{2}-1\right]-\ln\mathrm{K}_{1/4}\left(\rho\left(1+k\right)^{2}\right)+\ln\mathrm{K}_{1/4}\left(\rho\right).
\end{align}
By direct differentiation we obtain the exact two point correlation
function
\begin{align}
m^{2}\left\langle q^{2}\right\rangle _{K} & =4\rho\left(1+k\right)\left[\frac{\mathrm{K}_{3/4}\left(\rho\left(1+k\right)^{2}\right)}{\mathrm{K}_{1/4}\left(\rho\left(1+k\right)^{2}\right)}-1\right],
\end{align}
or for the original ($K=0$) theory,
\begin{align}
m^{2}\bar{G} & =4\rho\left(\frac{\mathrm{K}_{3/4}\left(\rho\right)}{\mathrm{K}_{1/4}\left(\rho\right)}-1\right).\label{eq:exact-G-soln}
\end{align}
Like $Z$, $\bar{G}$ possesses a branch cut discontinuity from $\lambda=0$
to $\lambda=-\infty$. At $\lambda=0$ one obtains the usual free
(Gaussian) theory result $\bar{G}=G_{0}=m^{-2}$. In the strong coupling
limit, $\lambda\to\infty$,
$\bar{G}\sim\left[2\sqrt{6}\Gamma\left(\frac{3}{4}\right)/\Gamma\left(\frac{1}{4}\right)\right] \lambda^{-1/2} + \mathcal{O}\left(\lambda^{-1}\right)$.
$\bar{G}$ is shown in Figure \ref{fig:Gexact}, from which the branch cut is obvious.
This can also be seen in more detail in the complex $\lambda$ plane
as shown in Figure \ref{fig:Gexact-analytic-behaviour}. One sees
that not only does $\bar{G}$ possess a branch cut, it is analytic
in the cut plane and is in fact a Herglotz-Nevanlinna function (i.e.
$\bar{G}\left(\lambda\right)^{\star}=\bar{G}\left(\lambda^{\star}\right)$
where $\star$ is complex conjugation). This means that $\bar{G}$
has a nice integral representation which we derive now in order to
quantify the branch cut.

\begin{figure}
\includegraphics[width=.8\columnwidth]{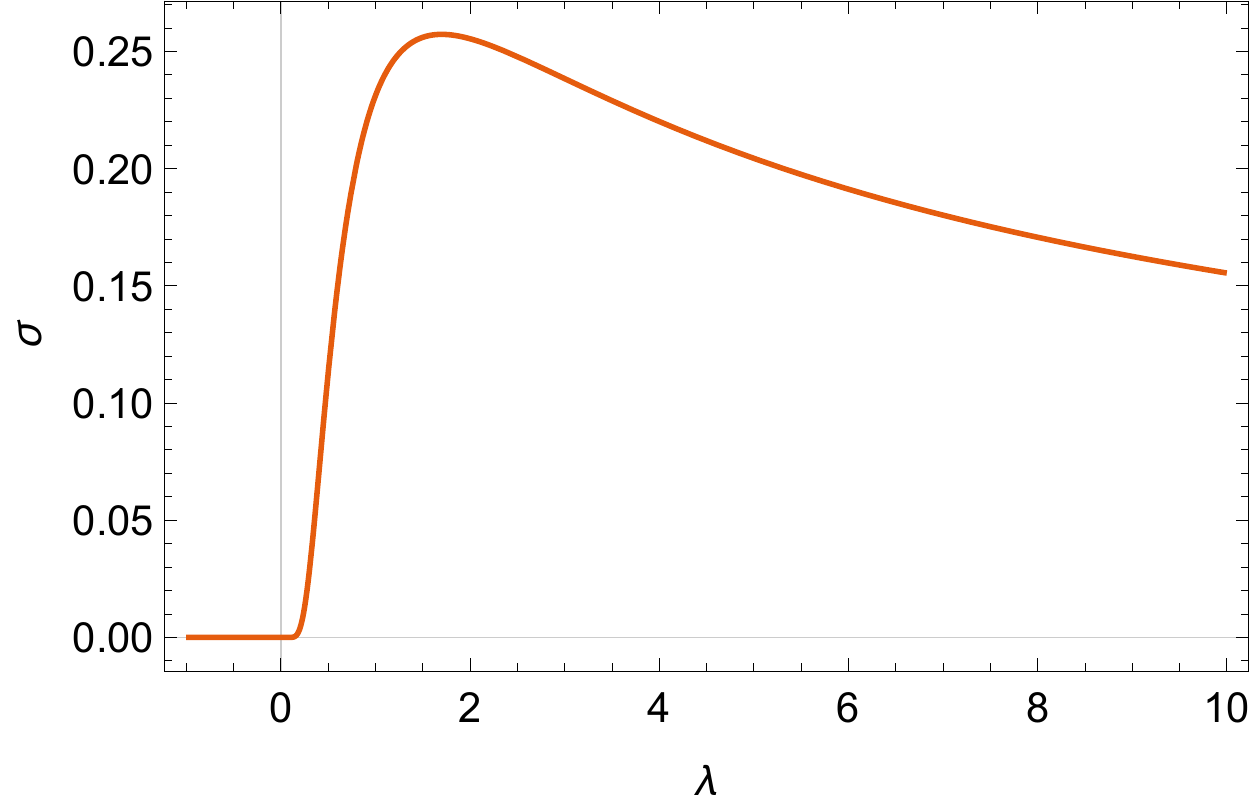}

\protect\caption{\label{fig:sigma-exact}Spectral function $\sigma\left(\lambda\right)$
of \eqref{eq:spectral-function-exact} for $m=1$.}
\end{figure}

We obtain the integral representation for $\bar{G}$ using the Cauchy formula
\begin{equation}
\bar{G}\left(\lambda\right)=\frac{1}{2\pi i}\oint_{C}\frac{\bar{G}\left(\lambda'\right)}{\lambda'-\lambda}\mathrm{d}\lambda',
\end{equation}
where the contour $C$ circles $\lambda$ in the counter-clockwise
direction and avoids the cut. Deforming the contour to run on the
circle at infinity and around the cut and using $\bar{G}\left(\lambda\right)\to0$
as $\left|\lambda\right|\to\infty$, we can write the integral in
terms of a spectral function $\sigma\left(s\right)=\left[\bar{G}\left(-s-i\epsilon\right)-\bar{G}\left(-s+i\epsilon\right)\right]\Theta\left(s\right)/2\pi i=\mathrm{Im}\left[\bar{G}\left(-s-i\epsilon\right)\right]\Theta\left(s\right)/\pi$
where $\Theta\left(s\right)$ is the Heaviside step function, such
that
\begin{equation}
\bar{G}\left(\lambda\right)=\int_{0}^{\infty}\mathrm{d}s\frac{\sigma\left(s\right)}{s+\lambda}.\label{eq:spectral-function-def}
\end{equation}
We find
\begin{equation}
\sigma\left(\lambda\right)=-\frac{4\sqrt{2}}{m^{2}\pi^{2}}\frac{1}{\mathrm{Im}\left[I_{-\frac{1}{4}}\left(-\rho\right)^{2}-I_{\frac{1}{4}}\left(-\rho\right)^{2}\right]}\Theta\left(\lambda\right),\label{eq:spectral-function-exact}
\end{equation}
where $I_{\pm\frac{1}{4}}\left(\rho\right)$ are modified Bessel functions
of the first kind.\footnote{Note that the physical interpretation of this spectral function is
unrelated to the usual one in field theory since, for one thing, there
is no such thing as energy in zero dimensions. We consider $\sigma\left(s\right)$
a purely formal device that gives information about the analytic structure
of $\bar{G}$.} $\sigma\left(\lambda\right)$ is shown in Figure \ref{fig:sigma-exact}.

\section{Perturbation theory and re-summations\label{sec:Perturbation-theory-and-resummations}}

\subsection{Perturbation theory\label{sub:Perturbation-theory}}

At small coupling one often uses perturbation theory in $\lambda$
which proceeds by expanding the exponential
\begin{align}
Z\left[K\right] & =N\int_{-\infty}^{\infty}\mathrm{d}q\exp\left(-\frac{1}{2}\left(m^{2}+K\right)q^{2}-\frac{1}{4!}\lambda q^{4}\right)\nonumber \\
 & =N\int_{-\infty}^{\infty}\mathrm{d}q\sum_{n=0}^{\infty}\frac{1}{n!}\left(-\frac{1}{4!}\lambda q^{4}\right)^{n}\exp\left(-\frac{1}{2}\left(m^{2}+K\right)q^{2}\right)\nonumber \\
 & \sim\sum_{n=0}^{\infty}\frac{1}{n!}\left(-\frac{1}{4!}\lambda\right)^{n}N\int_{-\infty}^{\infty}\mathrm{d}qq^{4n}\exp\left(-\frac{1}{2}\left(m^{2}+K\right)q^{2}\right)\nonumber \\
 & =\sum_{n=0}^{\infty}\frac{1}{n!}\left(-\frac{1}{4!}\lambda\right)^{n}N2^{2n+\frac{1}{2}}\Gamma\left(2n+\frac{1}{2}\right)\frac{1}{\left(m^{2}+K\right)^{2n+\frac{1}{2}}}.
\end{align}
In the third line we have formally interchanged the sum and integral,
leading to an asymptotic rather than convergent series for $Z\left[K\right]$.
Again, $N$ is determined by $Z\left[0\right]=1$, giving
\begin{equation}
N=\left[\sum_{n=0}^{\infty}\frac{1}{n!}\left(-\frac{1}{4!}\lambda\right)^{n}2^{2n+\frac{1}{2}}\Gamma\left(2n+\frac{1}{2}\right)\frac{1}{\left(m^{2}\right)^{2n+\frac{1}{2}}}\right]^{-1},
\end{equation}
so
\begin{equation}
Z\left[K\right]=\frac{\sum_{n=0}^{\infty}\frac{1}{n!}\left(-\frac{1}{4!}\lambda\right)^{n}2^{2n}\Gamma\left(2n+\frac{1}{2}\right)\frac{1}{\left(m^{2}+K\right)^{2n+\frac{1}{2}}}}{\sum_{n=0}^{\infty}\frac{1}{n!}\left(-\frac{1}{4!}\lambda\right)^{n}2^{2n}\Gamma\left(2n+\frac{1}{2}\right)\frac{1}{\left(m^{2}\right)^{2n+\frac{1}{2}}}},
\end{equation}
and
\begin{equation}
W\left[K\right]=-\ln\sum_{n=0}^{\infty}\frac{1}{n!}\left(-\frac{1}{4!}\lambda\right)^{n}2^{2n}\Gamma\left(2n+\frac{1}{2}\right)\frac{1}{\left(m^{2}+K\right)^{2n+\frac{1}{2}}}+\text{const}.
\end{equation}

\begin{figure}
\includegraphics[width=1\columnwidth]{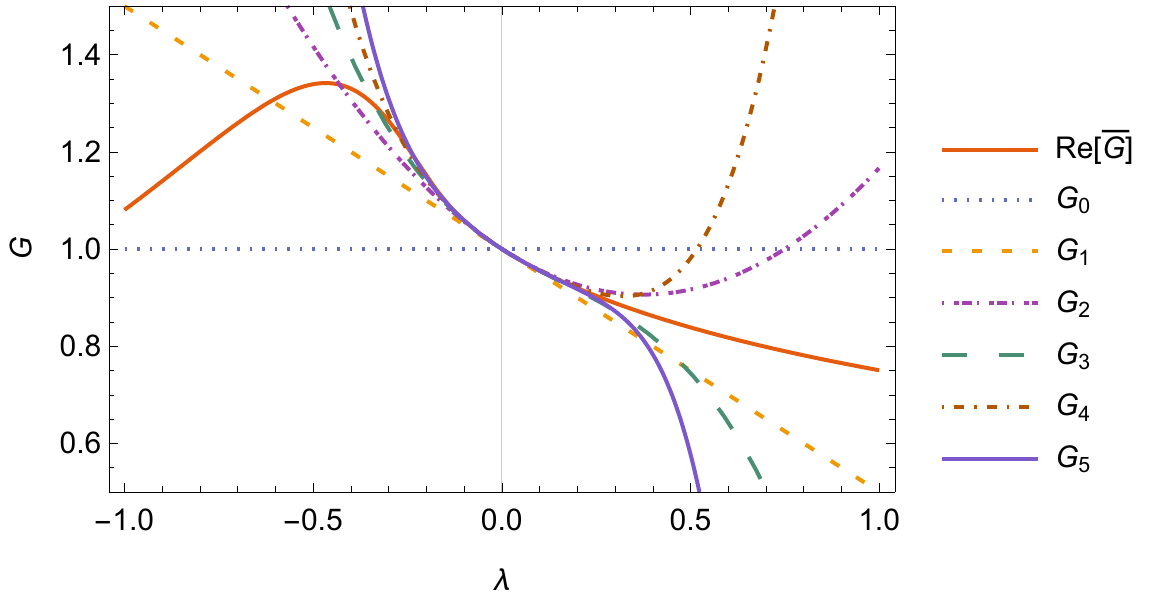}

\protect\caption{\label{fig:G-pert-series}Perturbative approximations to $\bar{G}$
up to $\mathcal{O}\left(\lambda^{5}\right)$ for $m=1$, compared
to the exact solution.}
\end{figure}

From this we find
\begin{align}
\bar{G} & =2\partial_{K}W\nonumber \\
 & =\frac{1}{m^{2}}+\frac{4}{m^{2}}\frac{\sum_{n=1}^{\infty}\frac{1}{\left(n-1\right)!}\left(-\frac{1}{4!}\lambda\right)^{n}2^{2n}\Gamma\left(2n+\frac{1}{2}\right)\frac{1}{m^{4n}}}{\sum_{n=0}^{\infty}\frac{1}{n!}\left(-\frac{1}{4!}\lambda\right)^{n}2^{2n}\Gamma\left(2n+\frac{1}{2}\right)\frac{1}{m^{4n}}}\nonumber \\
 & =\frac{1}{m^{2}}-\frac{\lambda}{2m^{6}}+\frac{2\lambda^{2}}{3m^{10}}-\frac{11\lambda^{3}}{8m^{14}}+\frac{34\lambda^{4}}{9m^{18}}+\cdots.\label{eq:G-perturbation-series}
\end{align}
The perturbative approximations $G_n$ are the $\mathcal{O}\left(\lambda^n\right)$
truncations of this series. The first few are shown compared to the exact $\bar{G}$
in Figure \ref{fig:G-pert-series}. Note that these are simply the
low order Taylor series approximations to $\bar{G}$. These approximations
apparently converge poorly to the exact solution, and in fact we will
shortly show that the series diverges.

The series for $\bar{G}$ can be described in terms of Feynman diagrams
by the following rules:
\begin{enumerate}
\item Draw all connected graphs with
two external lines (i.e. lines with one end not connected to any vertex)
constructed from lines and four point vertices.
\item Associate to each line a factor $G_{0}=1/m^{2}$.
\item Associate to each vertex a factor $-\lambda$.
\item Divide by an overall symmetry factor being the order of the symmetry
group of the diagram under permutations of lines and vertices.
\end{enumerate}
We illustrate these rules by giving the first few terms in $\bar{G}$:
\begin{fmffile}{perturbation-theory-propagator}
\begin{align}
\bar{G}&=\parbox{15mm}{\begin{fmfgraph}(15,15)
\fmfleft{i}\fmfright{o}\fmf{plain}{i,o}\end{fmfgraph}}
+\frac{1}{2}\parbox{15mm}{\begin{fmfgraph}(15,15)
\fmfleft{i}\fmfright{o}\fmf{plain}{i,v,v,o}
\fmfdot{v}\end{fmfgraph}}
+\frac{1}{4}\parbox{20mm}{\begin{fmfgraph}(20,15)
\fmfleft{i}\fmfright{o}\fmf{plain}{i,v1,v1,v2,v2,o}
\fmfdot{v1}\fmfdot{v2}\end{fmfgraph}}
+\frac{1}{4}\parbox{20mm}{\begin{fmfgraph}(20,15)
\fmfleft{i}\fmfright{o}\fmftop{t}
\fmf{plain}{i,v1,o}\fmffreeze
\fmf{phantom}{v1,v2}
\fmf{phantom}{v2,t}\fmffreeze
\fmf{plain,left}{v1,v2,t,v2,v1}
\fmfdot{v1,v2}\end{fmfgraph}}
+\frac{1}{6}\parbox{20mm}{\begin{fmfgraph}(20,15)
\fmfleft{i}\fmfright{o}
\fmf{plain}{i,v1,v2,o}
\fmf{plain,left,tension=0}{v1,v2,v1}
\fmfdot{v1,v2}\end{fmfgraph}}\nonumber\\
& + \frac{1}{8}\parbox{25mm}{
\begin{fmfgraph}(25,15)
\fmfleft{i}\fmfright{o}
\fmf{plain}{i,v1,v1,v2,v2,v3,v3,o}
\fmfdot{v1,v2,v3}
\end{fmfgraph}}
+\frac{1}{8}\parbox{25mm}{
\begin{fmfgraph}(25,15)
\fmfleft{i}\fmfright{o}\fmftop{tl,t1,t2,tr}
\fmf{plain}{i,v1,v2,o}\fmffreeze
\fmf{phantom}{v1,x1,t1}\fmffreeze
\fmf{phantom}{v2,v3,t2}\fmffreeze
\fmf{plain,left}{v1,x1,v1}
\fmf{plain,left}{v2,v3,v2}
\fmf{plain,left}{v3,t2,v3}
\fmfdot{v1,v2,v3}
\end{fmfgraph}}
+\frac{1}{8}\parbox{25mm}{
\begin{fmfgraph}(25,15)
\fmfleft{i}\fmfright{o}\fmftop{tl,t2,t1,tr}
\fmf{plain}{i,v2,v1,o}\fmffreeze
\fmf{phantom}{v1,x1,t1}\fmffreeze
\fmf{phantom}{v2,v3,t2}\fmffreeze
\fmf{plain,left}{v1,x1,v1}
\fmf{plain,left}{v2,v3,v2}
\fmf{plain,left}{v3,t2,v3}
\fmfdot{v1,v2,v3}
\end{fmfgraph}}\nonumber\\
 & +\frac{1}{12}\parbox{25mm}{
\begin{fmfgraph}(25,20)
\fmfleft{i}\fmfright{o}
\fmf{plain}{i,v1,v1,v2,v3,o}
\fmf{plain,left,tension=0}{v2,v3,v2}
\fmfdot{v1,v2,v3}
\end{fmfgraph}}
+\frac{1}{12}\parbox{25mm}{
\begin{fmfgraph}(25,20)
\fmfleft{i}\fmfright{o}
\fmf{plain}{i,v2,v3,v1,v1,o}
\fmf{plain,left,tension=0}{v2,v3,v2}
\fmfdot{v1,v2,v3}
\end{fmfgraph}}
+\frac{1}{8}\parbox{25mm}{
\begin{fmfgraph}(25,20)
\fmfleft{i}\fmfright{o}\fmftop{t}
\fmf{plain}{i,v1,o}\fmffreeze
\fmf{phantom}{v1,v2,v3,t}\fmffreeze
\fmf{plain,left}{v1,v2,v3,t,v3,v2,v1}
\fmfdot{v1,v2,v3}
\end{fmfgraph}}\nonumber\\
 & +\frac{1}{12}\parbox{25mm}{
\begin{fmfgraph}(25,25)
\fmfleft{i}\fmfright{o}\fmftop{tl,t1,t2,tr}
\fmf{plain}{i,x1,v1,x2,o}\fmffreeze
\fmf{phantom}{x1,v2,t1}\fmffreeze
\fmf{phantom}{x2,v3,t2}\fmffreeze
\fmf{plain,left=0.5}{v1,v2}
\fmf{plain,right=0.5}{v1,v3}
\fmf{plain,left=0.5}{v2,v3,v2}
\fmf{plain}{v2,v3}
\fmfdot{v1,v2,v3}
\end{fmfgraph}}
+\frac{1}{4}\parbox{25mm}{
\begin{fmfgraph}(25,25)
\fmfleft{i}\fmfright{o}\fmftop{t}
\fmf{plain}{i,v1,v2,o}\fmffreeze
\fmf{plain,left=0.5}{v1,v3,v2}
\fmf{plain,right}{v1,v2}
\fmf{phantom,tension=0.1}{v3,t}
\fmf{plain,left}{v3,t,v3}
\fmfdot{v1,v2,v3}
\end{fmfgraph}}
+\frac{1}{8}\parbox{25mm}{
\begin{fmfgraph}(25,25)
\fmfleft{lm2,lm1,i,l1,l2}\fmfright{rm2,rm1,o,r1,r2}\fmftop{tl,t1,t2,t3,tr}
\fmf{plain}{i,v1,o}\fmffreeze
\fmf{phantom}{l1,v2,v3,r1}\fmffreeze
\fmf{plain,left=0.6}{v1,v2}
\fmf{plain,left=0.4}{v2,v3}
\fmf{plain,left=0.6}{v3,v1}
\fmf{plain,left}{v2,t1,v2}\fmf{plain,left}{v3,t3,v3}
\fmfdot{v1,v2,v3}
\end{fmfgraph}}\nonumber\\
 & +\frac{1}{4}\parbox{25mm}{
\begin{fmfgraph}(25,15)
\fmfleft{i}\fmfright{o}\fmftop{t}
\fmf{plain}{i,v1,x1,v2,o}\fmffreeze
\fmf{phantom}{t,v3,x1}\fmffreeze
\fmf{plain,right=0.3}{v1,v3,v2}
\fmf{plain,left=0.5}{v1,v3,v2}
\fmfdot{v1,v2,v3}
\end{fmfgraph}}
+\mathcal{O}\left(\lambda^4\right).\label{eq:G-perturbation-series-diagrams}
\end{align}
\end{fmffile}
$W\left[K\right]$ can be written as a similar diagrammatic series
in terms of connected vacuum diagrams (i.e., those with no external
lines).

\begin{figure}
\includegraphics[width=.75\columnwidth]{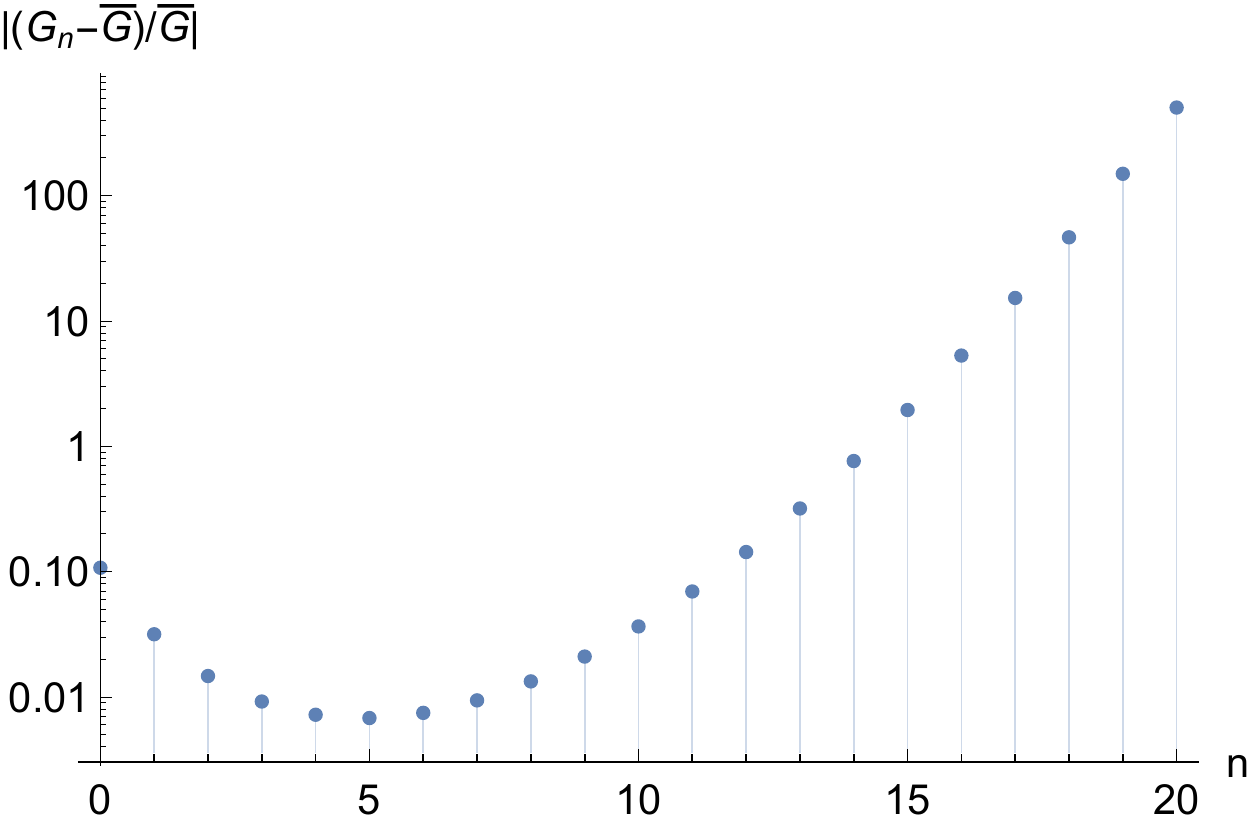}

\protect\caption{\label{fig:Gpert-asymptotic-series}Relative error $\left|\left(G_{n}-\bar{G}\right)/\bar{G}\right|$of
the $n$-th perturbative approximation to $\bar{G}$ for $m=1$ and
$\lambda=1/4$, showing the decreasing then increasing behaviour typical
of asymptotic series.}
\end{figure}

The series \eqref{eq:G-perturbation-series} has the form $\bar{G}=m^{-2}\sum_{n=0}^{\infty}c_{n}\left(\lambda/m^{4}\right)^{n}$
where the coefficients asymptotically obey $c_{n+1}\sim-\frac{2}{3}nc_{n}$
as $n\to\infty$, thus the radius of convergence of the series is
zero. This is consistent with the fact that we are perturbing around
a branch point of the exact solution: no approximation of $\bar{G}$
in terms of analytic functions can converge at $\lambda=0$ because
$Z\left[K\right]$ is itself undefined for $\mathrm{Re}\lambda<0$. The terms
of the series start to increase when $c_{n+1}\left(\lambda/m^{4}\right)\sim c_{n}$,
i.e. $n\sim3m^{4}/2\lambda$, meaning the series is useful for $\lambda\ll m^{4}$
but fails immediately for a moderately strong coupling $\lambda\approx m^{4}$.
This is typical asymptotic series behaviour as shown in Figure \ref{fig:Gpert-asymptotic-series}.
Extrapolating perturbation theory to strong coupling $\lambda\gg m^{4}$
is simply impossible, although the exact solution is well behaved
there. (In fact $\bar{G}$ can be expanded as $\bar{G}=m^{-2}\sum_{n=1}^{\infty}\tilde{c}_{n}\left(\lambda/m^{4}\right)^{-n/2}$,
displaying explicitly the branch point at $\lambda=\infty$.)

\subsection{Borel summation\label{sub:Borel-summation}}

We have seen that the series expansion for $\bar{G}$ diverges for
all $\lambda\ne0$. This is typical of perturbation series and usually
signals some singularity of the exact solution for unphysical values
of $\lambda$. In our case, indeed, the theory does not exist for
$\mathrm{Re}\lambda<0$ and the exact solution possesses a branch cut on the
negative $\lambda$ axis, a feature which cannot be reproduced in
any order of perturbation theory. However, the perturbation series
is asymptotic and does contain true information about the exact solution,
even if $\lambda$ is large enough that the series is not useful practically.
Because of the ubiquity of this phenomenon, mathematicians have invented
a number of series summation techniques which assign a finite value
to certain types of divergent series and which obey certain consistency
properties (e.g. the value assigned to a convergent series is just
its na\"{i}ve sum). Here we investigate Borel summation, which is capable
of summing factorially divergent series like \eqref{eq:G-perturbation-series}.

Suppose that $\sum_{n=0}^{\infty}a_{n}$ is a divergent series but
that the \emph{Borel transform} of the series, defined as 
\begin{equation}
\phi\left(x\right)=\sum_{n=0}^{\infty}\frac{a_{n}x^{n}}{n!},
\end{equation}
converges for sufficiently small $x$. Then, if the integral
\begin{equation}
B\left(x\right)=\int_{0}^{\infty}\mathrm{e}^{-t}\phi\left(xt\right)\mathrm{d}t\label{eq:borel-sum}
\end{equation}
exists the \emph{Borel sum} \citep{Bender1999,Ellis1996,Dorigoni2014}
of the divergent series is defined as $\mathcal{B}\left[\sum_{n=0}^{\infty}a_{n}\right]\equiv B\left(1\right)$.
This definition is justified by substituting the series for $\phi\left(xt\right)$
into the integral and evaluating term-wise and noting that $B\left(x\right)\sim\sum_{n=0}^{\infty}a_{n}x^{n}$.
The main drawback of Borel summation is that one must know the precise
form of $a_{n}$ for all $n$ to compute $\phi\left(x\right)$, which is rarely the
case in field theory. For this reason Borel summation cannot be usefully applied
directly. However, one may use Pad\'e approximants as discussed in the next
section to recast the Borel transform in a useful way.

We note that the key to Borel-summability of the perturbation series
is the alternating sign $\left(-1\right)^{n}$ of the $n$-th order
term. To see this consider the two series
\begin{align}
S_{1} & =\sum_{n=0}^{\infty}\left(-\lambda\right)^{n}n!,\\
S_{2} & =\sum_{n=0}^{\infty}\lambda^{n}n!,
\end{align}
which differ only by the alternating sign. The Borel transforms $\phi_{1,2}\left(x\right)$
are
\begin{align}
\phi_{1}\left(x\right) & =\sum_{n=0}^{\infty}\left(-\lambda x\right)^{n}=\frac{1}{1+\lambda x},\\
\phi_{2}\left(x\right) & =\sum_{n=0}^{\infty}\left(\lambda x\right)^{n}=\frac{1}{1-\lambda x},
\end{align}
and the Borel sums are 
\begin{align}
\mathcal{B}\left[S_{1}\right]=B_{1}\left(1\right) & =\int_{0}^{\infty}\frac{\mathrm{e}^{-t}}{1+\lambda t}\mathrm{d}t,\\
\mathcal{B}\left[S_{2}\right]=B_{2}\left(1\right) & =\int_{0}^{\infty}\frac{\mathrm{e}^{-t}}{1-\lambda t}\mathrm{d}t.
\end{align}
In the first case the integral exists and $\mathcal{B}\left[S_{1}\right]=\lambda^{-1}\mathrm{e}^{1/\lambda}\Gamma\left(0,\frac{1}{\lambda}\right)$
where $\Gamma\left(a,b\right)=\int_{b}^{\infty}t^{a-1}\mathrm{e}^{-t}\mathrm{d}t$
is the incomplete gamma function. However, the second integral hits
a pole at $t=1/\lambda$. There is no natural prescription for avoiding
the pole, which leads to an ambiguity in the sum of $\pm\pi i\lambda^{-1}\mathrm{e}^{-1/\lambda}$.
This is a non-perturbative ambiguity called a \emph{renormalon} \citep{Beneke1999}. In
every known case where this arises in field theory the renormalon
is connected to a non-perturbative finite action solution of the field
equations, i.e. an instanton or soliton, and a correct evaluation
of the path integral which sums over \emph{all} saddle points (not
just perturbative ones) removes the ambiguity. Key to the practical
application of Borel summation is the location and classification
of all renormalons in a given theory \citep{Crutchfield1979}. Sophisticated
techniques have been developed to deal with this situation which are
beyond the scope of this paper \citep{Dorigoni2014,Dunne2014,Basar2013}.

\subsection{Pad\'e approximation\label{sub:Pade-approximation}}

Borel summation on its own has limited usefulness in practice because one often
only knows a few low order terms of perturbation theory, and the potential
existence of renormalon singularities. There exists another technique
which often improves perturbation series and is far more useful in
practice (and is often combined with Borel summation). Pad\'e approximation
approximates a function by rational polynomials which generally converge
rapidly, are very useful for numerical computation and help to estimate
the location of singularities of the function in the complex plane.
Many software packages include standard routines for evaluating Pad\'e approximants,
for instance the \noun{PadeApproximant} function in \noun{Mathematica}.
In this section we apply Pad\'e approximants directly to the Green function
and then to the Borel transform. We find that the latter approach is clearly the better one.

The $\left(N,M\right)$-Pad\'e approximant of a function $\sum_{n=0}^{\infty}a_{n}x^{n}$
is given by \citep{Bender1999}
\begin{equation}
P_{M}^{N}\left(x\right)=\frac{\sum_{n=0}^{N}A_{n}x^{n}}{\sum_{n=0}^{M}B_{n}x^{n}},
\end{equation}
where without loss of generality one takes $B_{0}=1$. The remaining
$N+M+1$ coefficients are chosen so that the Taylor series of $P_{M}^{N}\left(x\right)$
matches the perturbation series up to $\mathcal{O}\left(x^{N+M}\right)$.
Due to the denominator, Pad\'e approximants develop poles in the complex
$x$-plane, allowing the close approximation of more singular functions
than Taylor series are capable of. Examples of the use of Pad\'e approximants
in field theory can be found in \citep{Ellis1996,Kraemmer2004} and
references therein.

\begin{figure}
\includegraphics[width=1\columnwidth]{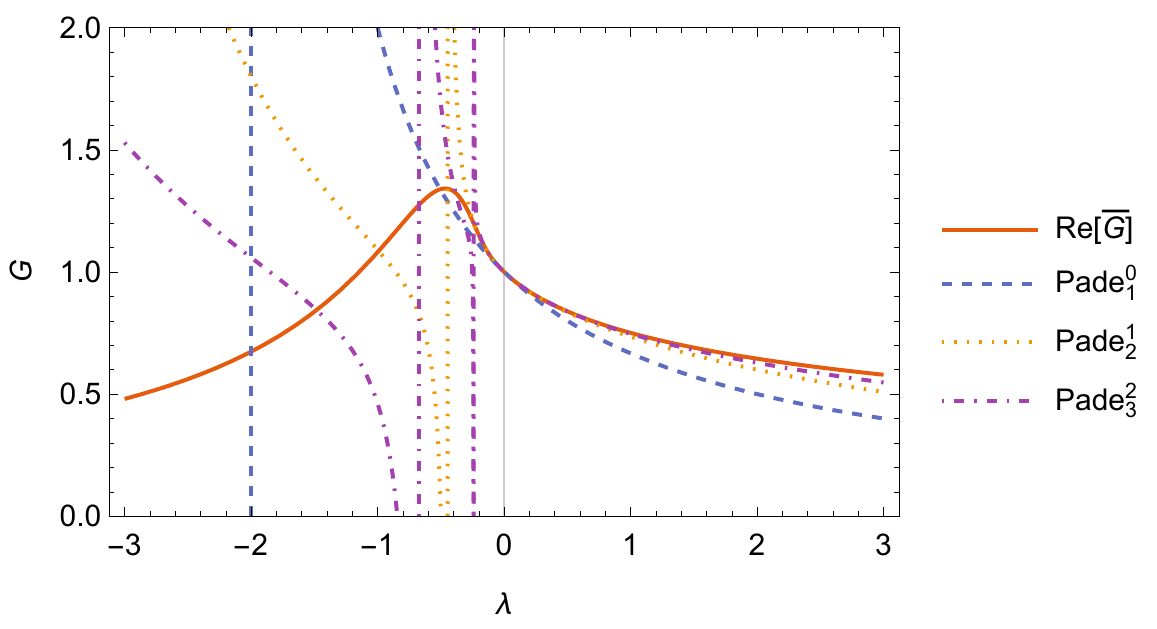}

\protect\caption{\label{fig:Pade-approx}First three Pad\'e approximants to $\bar{G}$.
Note the simple poles developed for $\lambda<0$.}
\end{figure}

Now we find the Pad\'e approximants to $m^{2}\bar{G}$ with $x=\lambda/m^{4}$.
Here we restrict attention to the approximants where $M=N+1$.
This guarantees that $P_{M}^{N}\to0$ as $\lambda\to\infty$. If we had used
the usual diagonal approximants ($N=M$) we would find an unphysical constant
term $P_{N}^{N}\to A_{N}/B_{N}$ as $\lambda\to\infty$. Note that
it is impossible to match the true $1/\sqrt{\lambda}$ behaviour of
$\bar{G}$ as $\lambda\to\infty$ using Pad\'e approximants centred
on the origin. The best that is possible in this limit is $\sim\lambda^{-1}$.
(As it happens, using $x\propto\sqrt{\lambda}$ does not allow one
to resolve this issue: the same approximants are found only with $x^{2}$
everywhere in place of $x$. This is because the $1/\sqrt{\lambda}$
behaviour is due to the branch point at infinity, which is infinitely
far from the origin where the Pad\'e approximants are matched to perturbation
theory. Low order Pad\'e approximants can extract information about
the branch point near the origin, but evidently not the one at infinity.)
The first five approximants for $m^{2}\bar{G}$ are shown in Table
\ref{tab:pade-approximants} and the first three are plotted in Figure
\ref{fig:Pade-approx} with comparison to the exact $\bar{G}$. Note
that the existence of the integral representation \eqref{eq:spectral-function-def}
for $\bar{G}$ implies that $\bar{G}$ is a \emph{Stieltjes} function,
meaning one can prove convergence properties for the Pad\'e approximants
as $N,M\to\infty$, though we are not concerned with this analysis
here (see \citep{Bender1999} for details).

\begin{table}[tbp]
\protect\caption{\label{tab:pade-approximants}First few Pad\'e approximants for $m^{2}\bar{G}$.}

\renewcommand{\arraystretch}{2.5}

\centering
\begin{tabular}{|l|l|}
\hline 
$N$ & $P_{N+1}^{N}$\\
\hline 
0 & $\begin{aligned}[t]\frac{1}{1+\frac{\lambda}{2m^{4}}}\end{aligned}$\\
1 & $\begin{aligned}[t]\frac{1+\frac{2\lambda}{m^{4}}}{1+\frac{5\lambda}{2m^{4}}+\frac{7\lambda^{2}}{12m^{8}}}\end{aligned}$\\
2 & $\begin{aligned}[t]\frac{1+\frac{16\lambda}{3m^{4}}+\frac{59\lambda^{2}}{12m^{8}}}{1+\frac{35\lambda}{6m^{4}}+\frac{43\lambda^{2}}{6m^{8}}+\frac{77\lambda^{3}}{72m^{12}}}\end{aligned}$\\
3 & $\begin{aligned}[t]\frac{1+\frac{10\lambda}{m^{4}}+\frac{155\lambda^{2}}{6m^{8}}+\frac{15\lambda^{3}}{m^{12}}}{1+\frac{21\lambda}{2m^{4}}+\frac{365\lambda^{2}}{12m^{8}}+\frac{295\lambda^{3}}{12m^{12}}+\frac{385\lambda^{4}}{144m^{16}}}\end{aligned}$\\
4 & $\begin{aligned}[t]\frac{1+\frac{16\lambda}{m^{4}}+\frac{315\lambda^{2}}{4m^{8}}+\frac{1190\lambda^{3}}{9m^{12}}+\frac{7945\lambda^{4}}{144m^{16}}}{1+\frac{33\lambda}{2m^{4}}+\frac{259\lambda^{2}}{3m^{8}}+\frac{11935\lambda^{3}}{72m^{12}}+\frac{14315\lambda^{4}}{144m^{16}}+\frac{7315\lambda^{5}}{864m^{20}}}\end{aligned}$\\
\hline 
\end{tabular}
\end{table}

\begin{figure}
\includegraphics[width=1\columnwidth]{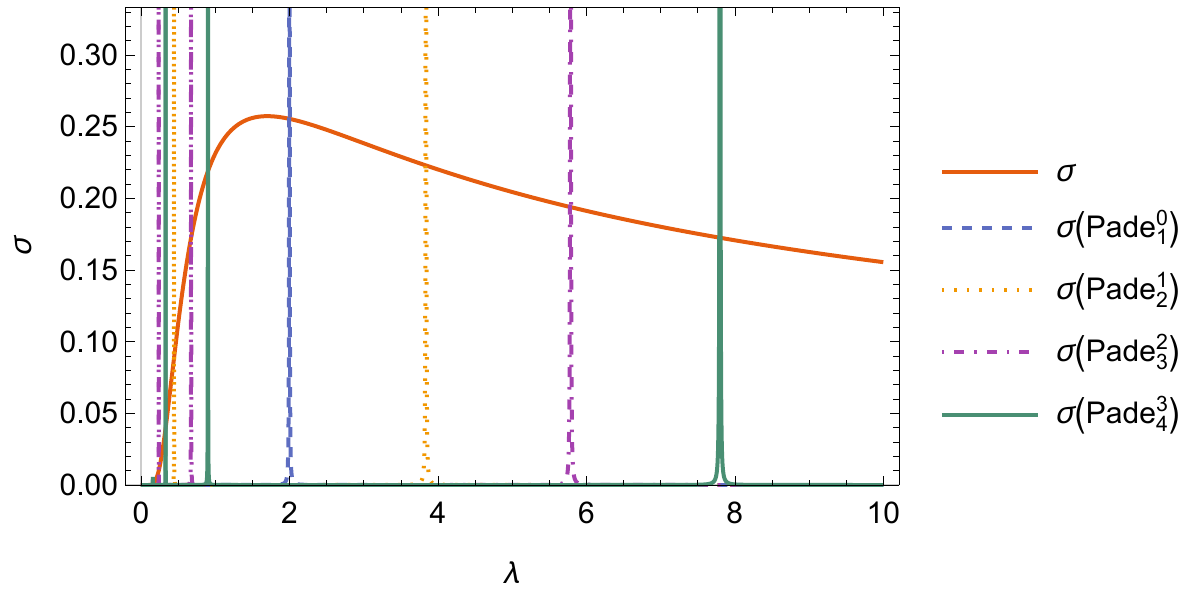}

\protect\caption{\label{fig:Pade-spectral-function}Pad\'e approximations to the spectral
function $\sigma\left(\lambda\right)$ consist of an increasingly
dense set of delta functions. (Note the delta functions have been
smoothed for visual purposes only.)}
\end{figure}

The $\left(N,N+1\right)$-Pad\'e approximant can also be written as
\begin{equation}
P_{N+1}^{N}=\sum_{i=0}^{N}\frac{r_{i}}{\lambda-p_{i}},
\end{equation}
where $r_{i}$ and $p_{i}$ are the $i$-th residue and pole respectively.
Note that since all of the coefficients in the denominators of Table
\ref{tab:pade-approximants} are positive and real, all of the poles
must either be on the negative real axis or they must be complex and
come in complex conjugate pairs. Numerical experiments suggest that
all the poles lie on the negative $\lambda$ axis, though we do not
know a proof of this for all $N$. Assuming this is generally true,
Pad\'e approximants give a representation of $\bar{G}$ which approximates
the continuous spectral function by a sum of delta functions
\begin{equation}
\sigma\left(s\right)\approx\sigma_{N}\left(s\right)\equiv\sum_{i=0}^{N}r_{i}\delta\left(s+p_{i}\right).
\end{equation}
As $N\to\infty$ the poles become denser and fill the negative $\lambda$
axis, eventually merging into a continuous branch cut. Similarly the
spectral function turns into a dense sum of delta functions which,
when considered acting on any sufficiently smooth test function, smooths
into a continuous function. The first few $\sigma_{N}$ are shown
next to the exact spectral function in Figure \ref{fig:Pade-spectral-function}
for comparison.

Now we consider Pad\'e approximation of the Borel transform of $\bar{G}$.
First we note the following connections between the Borel transform $\phi$ and $\bar{G}$ and $\sigma$:
\begin{equation}
v^{-1}\sigma\left(\frac{\lambda}{v}\right)\xleftarrow[x\to v]{\mathcal{L}^{-1}}\phi\left(x\right)\xrightarrow[x\to s]{\mathcal{L}}s^{-1}\bar{G}\left(\frac{\lambda}{s}\right).
\end{equation}
That is, the Green function is related to the Laplace transform of the Borel transform, while the spectral function is related to the inverse Laplace transform of the Borel transform. These relations can be shown using the definitions of $\phi$ and $\sigma$
and the integral representation of the (inverse) Laplace transform.
This allows us to extract the spectral function directly from the Borel transform. Note that each pole of $\phi$ yields by the inverse Laplace transform a term of the form $\lambda^{-1}\exp{\left(-k/\lambda\right)}$ in $\sigma$, where $k$ is controlled by the location of the pole. The general Borel-Pad\'e approximation for $\sigma$ is a superposition of terms of this form.

\begin{figure}
\includegraphics[width=1\columnwidth]{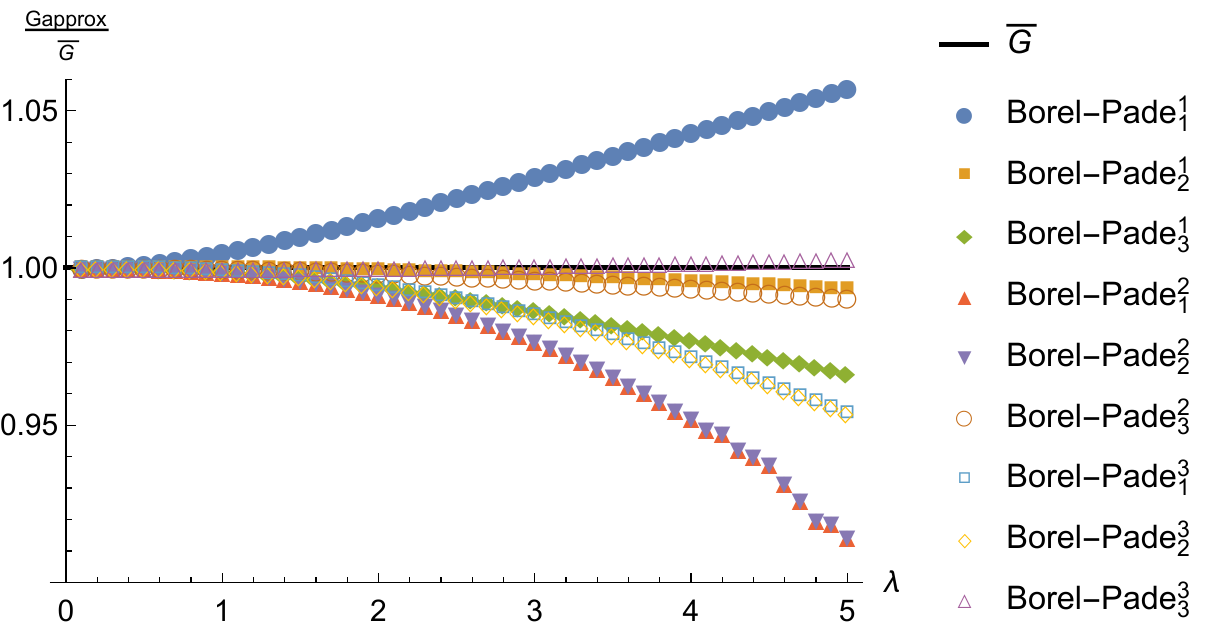}

\protect\caption{\label{fig:Gbar-PadeBorel}Borel-Pad\'e approximations to the Green
function $\bar{G}\left(\lambda\right)$ (ratio of approximant / exact). Approximants are calculated by numerical integration of \eqref{eq:borel-sum}.}
\end{figure}

\begin{figure}
\includegraphics[width=1\columnwidth]{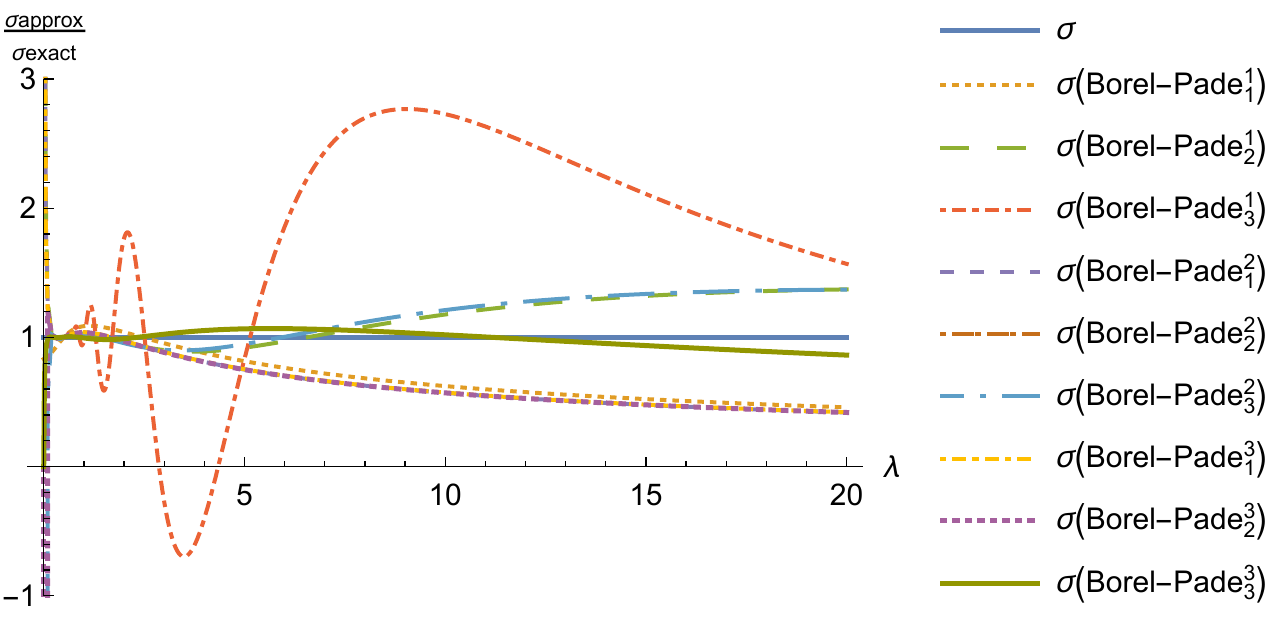}

\protect\caption{\label{fig:sigma-PadeBorel}Borel-Pad\'e approximations to the spectral
function $\sigma\left(\lambda\right)$ (ratio of approximant / exact). Note that the (2,2) approximant is off scale.}
\end{figure}

We show the first few approximants to $\bar{G}$ and $\sigma$ in Figures \ref{fig:Gbar-PadeBorel} and \ref{fig:sigma-PadeBorel} respectively. One sees that the low order Borel-Pad\'e Green functions are reasonably accurate (within a few percent for $\lambda\leq 5 m^4$) but the spectral functions are not approximated particularly well. Certain approximants to $\sigma$ oscillate erratically and even become negative for certain values of $\lambda$. Except for the fact that the best approximant is the highest order one plotted, there is no clear sense in which the Borel-Pad\'e approximations appear to converge to $\sigma$. However, even this bad approximation is at least a continuous function, as opposed to a sum of delta functions.

\section{2PI Approximations\label{sec:2PI-Approximations}}

The 2PI effective action is a functional of the Green function $G\equiv\left\langle q^{2}\right\rangle _{K}$
which is defined by the Legendre transform \citep{Berges2004,Cornwall1974}
\begin{equation}
\Gamma\left[G\right]=W\left[K\right]-K\partial_{K}W\left[K\right]
\end{equation}
where $K$ is solved for in terms of $G$. $\Gamma\left[G\right]$
obeys the equation of motion
\begin{equation}
\partial_{G}\Gamma\left[G\right]=-\frac{1}{2}K.
\end{equation}

The standard derivation of the 2PI action \citep{Berges2004,Cornwall1974}
gives in this case
\begin{equation}
\Gamma\left[G\right]=\frac{1}{2}\ln\left(G^{-1}\right)+\frac{1}{2}m^{2}G+\gamma_{2\mathrm{PI}}+\mathrm{const.},
\end{equation}
where $-\gamma_{2\mathrm{PI}}$ is (minus due to Euclidean conventions)
the sum of two particle irreducible vacuum graphs, i.e. those graphs
which do not fall apart when any two lines are cut, where the lines
are given by $G$ and vertices by $-\lambda$. Explicitly:\begin{fmffile}{twopi-diagrams}
\begin{align}
-\gamma_{2\text{PI}} &=
\frac{1}{8}\ \parbox[c]{15mm}{\begin{fmfgraph}(15,20)
\fmfleft{l}\fmfright{r}
\fmf{plain,left}{l,v,r,v,l}
\fmfdot{v}
\end{fmfgraph}}
\ +\frac{1}{4!2}\ \parbox{10mm}{\begin{fmfgraph}(10,20)
\fmfleft{l}\fmfright{r}
\fmf{plain,left}{l,r,l}
\fmf{plain,left=0.5}{l,r,l}
\fmfdot{l,r}
\end{fmfgraph}}
\ +\frac{1}{2^3 3!}\parbox{10mm}{\begin{fmfgraph}(10,10)
\fmfsurroundn{v}{3}
\fmfdotn{v}{3}
\fmfcyclen{plain,right=0.25}{v}{3}
\fmfrcyclen{plain,right=0.25}{v}{3}
\end{fmfgraph}}
+\mathcal{O}\left(\lambda^4\right).
\end{align}
\end{fmffile}

The equation of motion in the absence of sources is
\begin{equation}
G^{-1}=m^{2}+2\partial_{G}\gamma_{2\mathrm{PI}},\label{eq:2pi-G-eom}
\end{equation}
which has the diagram expansion
\begin{fmffile}{2pi-eom}
\begin{align}
G^{-1}=m^{2}-\frac{1}{2}\ \parbox{15mm}{\begin{fmfgraph}(15,15)
\fmfleft{i}\fmfright{o}\fmf{plain}{i,v,v,o}
\fmfdot{v}\end{fmfgraph}}
\ -\frac{1}{6}\ \parbox{20mm}{\begin{fmfgraph}(20,15)
\fmfleft{i}\fmfright{o}
\fmf{plain}{i,v1,v2,o}
\fmf{plain,left,tension=0}{v1,v2,v1}
\fmfdot{v1,v2}\end{fmfgraph}}
\ -\frac{1}{4}\parbox{25mm}{
\begin{fmfgraph}(25,20)
\fmfleft{i}\fmfright{o}\fmftop{t}
\fmf{plain}{i,v1,x1,v2,o}\fmffreeze
\fmf{phantom}{t,v3,x1}\fmffreeze
\fmf{plain,right=0.3}{v1,v3,v2}
\fmf{plain,left=0.5}{v1,v3,v2}
\fmfdot{v1,v2,v3}
\end{fmfgraph}}
+\mathcal{O}\left(\lambda^4\right).
\end{align}
\end{fmffile}
Notice that there is a dramatic reduction in the number of graphs
of a given order compared to perturbation theory \eqref{eq:G-perturbation-series-diagrams} due to the two-particle
irreducibility. This is one of the major benefits of 2PI approximation
schemes in realistic theories, such as gauge theories, where the Feynman
diagrams proliferate rapidly.

The equation of motion is Dyson's equation and the second term on the right hand side,
$-\Sigma=2\partial_{G}\gamma_{2\mathrm{PI}}$, represents the exact
one-particle irreducible self-energy of the propagator $G$. This
can be put into the usual form of a Dyson equation by noting $m^{2}=G_{0}^{-1}$
and multiplying both sides by $GG_{0}$:
\begin{equation}
G=G_{0}+G_{0}\Sigma G.\label{eq:2pi-Dyson-eq}
\end{equation}
Iterating this gives the infinite series 
\begin{equation}
G=G_{0}+G_{0}\Sigma G_{0}+G_{0}\Sigma G_{0}\Sigma G_{0}+\cdots.
\end{equation}
The main difference between this and the usual perturbative Dyson
equation is that $\Sigma$ contains exact propagators $G$ rather
than $G_{0}$. Inserting the expression above for $G$ into $\Sigma$
one finds that, even if one retains only a finite number of 2PI diagrams
in $\gamma_{2\text{PI}}$, $\Sigma$ contains an infinite series of
perturbative self-energy graphs. This is the motivation in the literature
for talking about 2PI as a resummation method.

By power counting (or counting line ends in the corresponding diagrams)
we find $\gamma_{2\mathrm{PI}}=\sum_{n=0}^{\infty}\gamma_{n}\left(\lambda G^{2}\right)^{n}$.
It is possible to derive the $\gamma_{n}$ by considering the symmetry
factors of the two particle irreducible Feynman diagrams, but we can
also obtain the $\gamma_{n}$ in a simple automated way using knowledge
of the exact solution $G=\bar{G}$. Substituting $\bar{G}$ into the
equation of motion and the expansion for $\gamma_{2\text{PI}}$, expanding
about $\lambda=0$ and matching powers of $\lambda$ we can determine
\begin{align}
\gamma_{2\mathrm{PI}} & =\frac{G^{2}\lambda}{8}-\frac{G^{4}\lambda^{2}}{48}+\frac{G^{6}\lambda^{3}}{48}-\frac{5G^{8}\lambda^{4}}{128}+\frac{101G^{10}\lambda^{5}}{960}-\frac{93G^{12}\lambda^{6}}{256}\nonumber \\
 & +\frac{8143G^{14}\lambda^{7}}{5376}-\frac{271217G^{16}\lambda^{8}}{36864}+\frac{374755G^{18}\lambda^{9}}{9216}-\frac{5151939G^{20}\lambda^{10}}{20480}\nonumber \\
 & +\frac{697775057G^{22}\lambda^{11}}{405504}-\frac{3802117511G^{24}\lambda^{12}}{294912}+\frac{201268707239G^{26}\lambda^{13}}{1916928}\nonumber \\
 & -\frac{11440081763125G^{28}\lambda^{14}}{12386304}+\frac{5148422676667G^{30}\lambda^{15}}{589824}-\frac{1665014342007385G^{32}\lambda^{16}}{18874368}\nonumber \\
 & +\frac{4231429245358235G^{34}\lambda^{17}}{4456448}-\frac{921138067678697395G^{36}\lambda^{18}}{84934656} +\mathcal{O}\left(\lambda^{19}G^{38}\right).\label{eq:gamma2pi-series}
\end{align}

We do not know of any closed form expression for either the coefficients
of this series or its sum (implicit analytic expressions can be derived;
however, these require the inversion of $G=\bar{G}\left(\lambda\right)$
for $\lambda\left(G\right)$, which is not known to us in closed form).
However, after the first few terms the coefficients seem to be well
approximated by $\gamma_{i+1}\sim-\frac{2}{3}i\gamma_{i}$, the same
as for the perturbation series. This has the hallmark of an asymptotic
series. Like perturbation theory, the 2PI series does not converge.

The first non-trivial contribution to the equation of motion gives
\begin{equation}
G_{\left(1\right)}^{-1}=m^{2}+\frac{\lambda}{2}G_{\left(1\right)},
\end{equation}
where the subscript $\left(1\right)$ indicates terms of order $\mathcal{O}\left(\lambda^{1}\right)$
have been kept. This has two solutions
\begin{equation}
G_{\left(1\right)}=\frac{-m^{2}\pm\sqrt{m^{4}+2\lambda}}{\lambda}.
\end{equation}
One of these solutions is unphysical and we must choose the $+$ sign.
As $\lambda\to0$, 
\begin{equation}
G_{\left(1\right)}\to\frac{1}{m^{2}}-\frac{\lambda}{2m^{6}}+\frac{\lambda^{2}}{2m^{10}}+\mathcal{O}\left(\lambda^{3}\right),
\end{equation}
which matches perturbation theory up to $\mathcal{O}\left(\lambda^{2}\right)$
terms as expected. However, unlike perturbation theory, the strong
coupling limit $\lambda\to\infty$ exists and gives
\begin{equation}
G_{\left(1\right)}\to\sqrt{\frac{2}{\lambda}}-\frac{m^{2}}{\lambda}+\frac{m^{4}}{\left(2\lambda\right)^{3/2}}+\mathcal{O}\left(\frac{1}{\lambda^{5/2}}\right).
\end{equation}
This series has the correct form in powers of $\lambda^{-1/2}$, though
the leading coefficient is incorrect by $\approx15\%$, and the sub-leading
coefficients are incorrect by $\approx40\%$, $70\%$ and $100\%$
etc. Nevertheless, it is remarkable to achieve any accuracy at all
given the simple nature of the approximation and the fact that $\gamma_{2\text{PI}}$
was truncated at leading order in $\lambda$! $G_{\left(1\right)}$
is a much more uniform approximation to $\bar{G}$ than the perturbative
approximation $m^{-2}-\lambda/2m^{6}$.

\begin{figure}
\includegraphics[width=1\columnwidth]{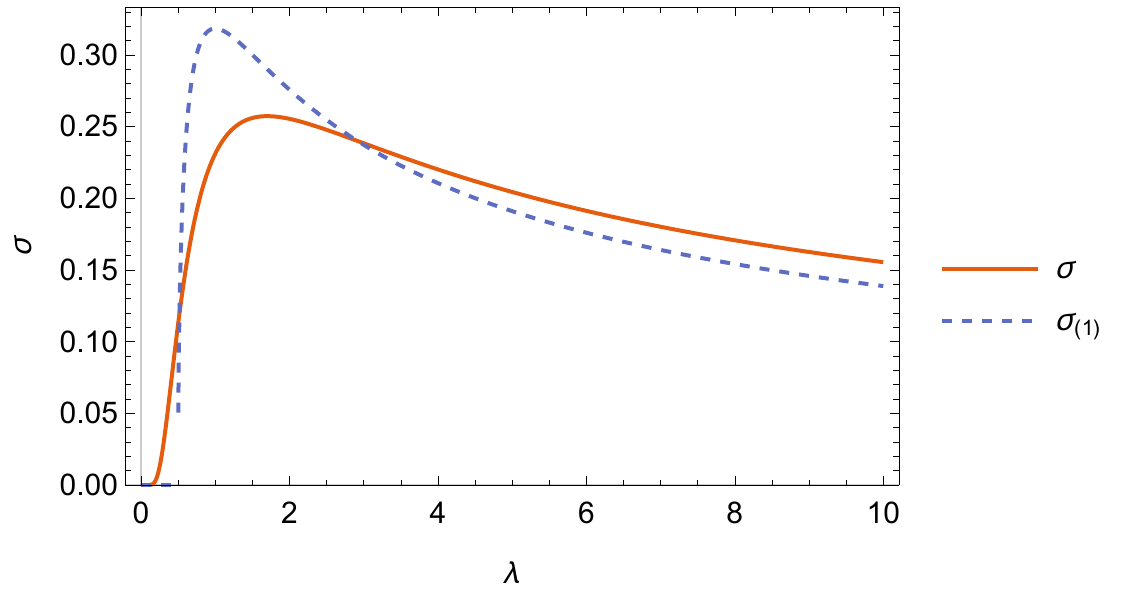}

\protect\caption{\label{fig:sigma-2pi-2loop}Comparison of exact spectral function
$\sigma\left(\lambda\right)$ \eqref{eq:spectral-function-exact}
with the two loop 2PI approximation $\sigma_{\left(1\right)}\left(\lambda\right)$
\eqref{eq:sigma-2loop-2pi-approx} for $m=1$. Already, the simplest
nontrivial 2PI truncation gives a much better approximation than perturbation
theory or Pad\'e approximants to arbitrary order.}
\end{figure}

This result is possible because of the branch cut $G_{\left(1\right)}$
possesses on the negative $\lambda$ axis. The discontinuity across
the cut
\begin{equation}
G_{\left(1\right)}\left(\lambda+i\epsilon\right)-G_{\left(1\right)}\left(\lambda-i\epsilon\right)=2i\frac{\sqrt{-m^{4}-2\lambda}}{\lambda}\Theta\left(-\lambda-\frac{m^{4}}{2}\right),
\end{equation}
gives the spectral function
\begin{equation}
\sigma_{\left(1\right)}\left(\lambda\right)=\frac{\sqrt{-m^{4}+2\lambda}}{\pi\lambda}\Theta\left(\lambda-\frac{m^{4}}{2}\right).\label{eq:sigma-2loop-2pi-approx}
\end{equation}
This is a far better approximation to the exact spectral function
than obtained from any of the other techniques, as can be seen from
Figure \ref{fig:sigma-2pi-2loop}.

\begin{figure}
\includegraphics[width=1\columnwidth]{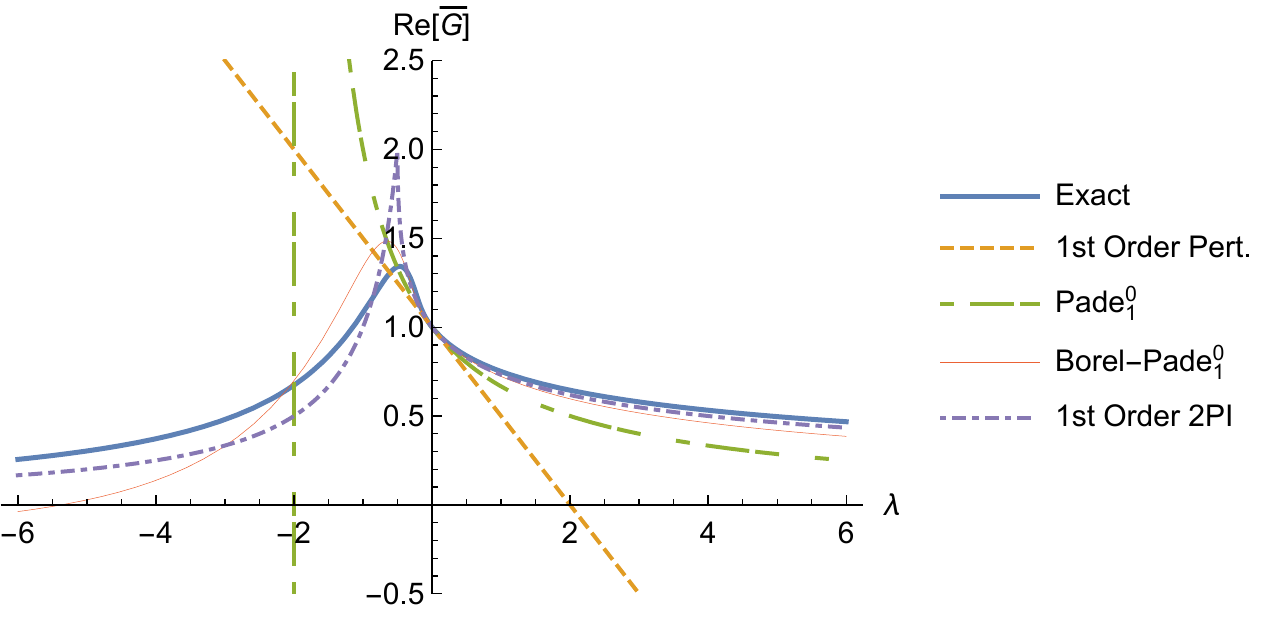}

\protect\caption{\label{fig:Gbar-all-first-order-approxs}Comparison of (the real part of)
the exact Green function $\bar{G}$ to each of the approximations discussed to first non-trivial order in each case.}
\end{figure}

We show all of the first order approximations in Figure \ref{fig:Gbar-all-first-order-approxs}.
Note that for $\lambda/m^4 \geq 0$ the best approximation is the 2PI, followed by
the Borel-Pad\'e, then by Pad\'e, then perturbation theory last of all.
The situation for negative $\lambda$ is complicated. The best approximation overall
is the 2PI, though it has an unphysical cusp where the branch cut starts
($\lambda/m^4 = -1/2$ in Figure \ref{fig:Gbar-all-first-order-approxs}).
It appears the 2PI approximation trades sensitivity to the exponentially small portion
of $\sigma$ in exchange for a better global approximation.
The Pad\'e approximation is good at small negative $\lambda$ but hits a pole at $\lambda/m^4 = -2$
and there loses all validity.
The Borel-Pad\'e approximation is smooth and more accurate than the 2PI near the peak of $\bar{G}$,
but eventually becomes invalid, even negative, at sufficiently large negative $\lambda$ ($\lambda/m^4 \lesssim -5.37$).

At $n$-th order, \eqref{eq:2pi-G-eom} is a degree $2n$ polynomial
in $G$ which has $2n$ roots, only one of which is physical. For
$n=2$ there are analytical expressions for the roots, though they
are very bulky, and for $n\geq3$ \eqref{eq:2pi-G-eom} must be solved
numerically. Picking out the correct root for a given value of $\lambda$
is tricky in general and we leave this exercise to the reader. However,
one can see on general grounds that truncations of \eqref{eq:2pi-G-eom}
give a singular perturbation theory in $\lambda$: at $\lambda=0$ the
physical solution starts at $G_{0}$ and the spurious solutions flow
in from infinity as inverse powers of $\lambda$ as $\lambda$ increases
to finite values.\footnote{We also note that at least one of the spurious solutions (and always
an odd number of them) are real since the coefficients in \eqref{eq:2pi-G-eom}
are real and complex solutions must occur in conjugate pairs.}
At strong coupling the roots generically approach each other and
one must carefully track them through the complex plane.
We suspect that a resummation of the 2PI series would remove some or all
of these spurious solutions, though we do not have a proof of this.
We examine two potential methods to achieve this in the next section and
find mixed results.

\section{Hybrid 2PI-Pad\'e\label{sec:Hybrid-2PI-Pade}}

Since the series of 2PI diagrams is asymptotic, we may consider using
a series summation method to improve convergence. Note that this is
logically independent of the resummations embodied in the 2PI approximation
itself. We may perform Pad\'e summation of the action term $\gamma_{2\text{PI}}$
or the equation of motion term $\partial_{G}\gamma_{2\text{PI}}$.
We consider both. First, consider Pad\'e summation of the action which
matches the series expansion up to order $N+M$:
\begin{equation}
\Gamma\left[G\right]=\frac{1}{2}\ln\left(G^{-1}\right)+\frac{1}{2}m^{2}G+\frac{\sum_{n=0}^{N}A_{n}^{\left(\gamma\right)}\left(\lambda G^{2}\right)^{n}}{\sum_{n=0}^{M}B_{n}^{\left(\gamma\right)}\left(\lambda G^{2}\right)^{n}}+\mathrm{const.}
\end{equation}
The equation of motion becomes
\begin{align}
G^{-1} & =m^{2}+2\frac{\mathrm{d}}{\mathrm{d}G}\frac{\sum_{n=0}^{N}A_{n}^{\left(\gamma\right)}\left(\lambda G^{2}\right)^{n}}{\sum_{n=0}^{M}B_{n}^{\left(\gamma\right)}\left(\lambda G^{2}\right)^{n}}\nonumber \\
 & =m^{2}+\frac{4\sum_{n=0}^{N}\sum_{k=0}^{M}\left(n-k\right)A_{n}^{\left(\gamma\right)}B_{k}^{\left(\gamma\right)}\lambda^{n+k}G^{2\left(n+k\right)-1}}{\left[\sum_{n=0}^{M}B_{n}^{\left(\gamma\right)}\left(\lambda G^{2}\right)^{n}\right]^{2}}.\label{eq:hybrid-attempt-1}
\end{align}
Multiplying by the denominator this becomes a polynomial equation
of degree $2\left(N+M\right)$ as expected. This equation will have
$2\left(N+M\right)-1$ spurious solutions as does the ordinary 2PI
approximation of matching order.

\begin{figure}
\includegraphics[width=1\columnwidth]{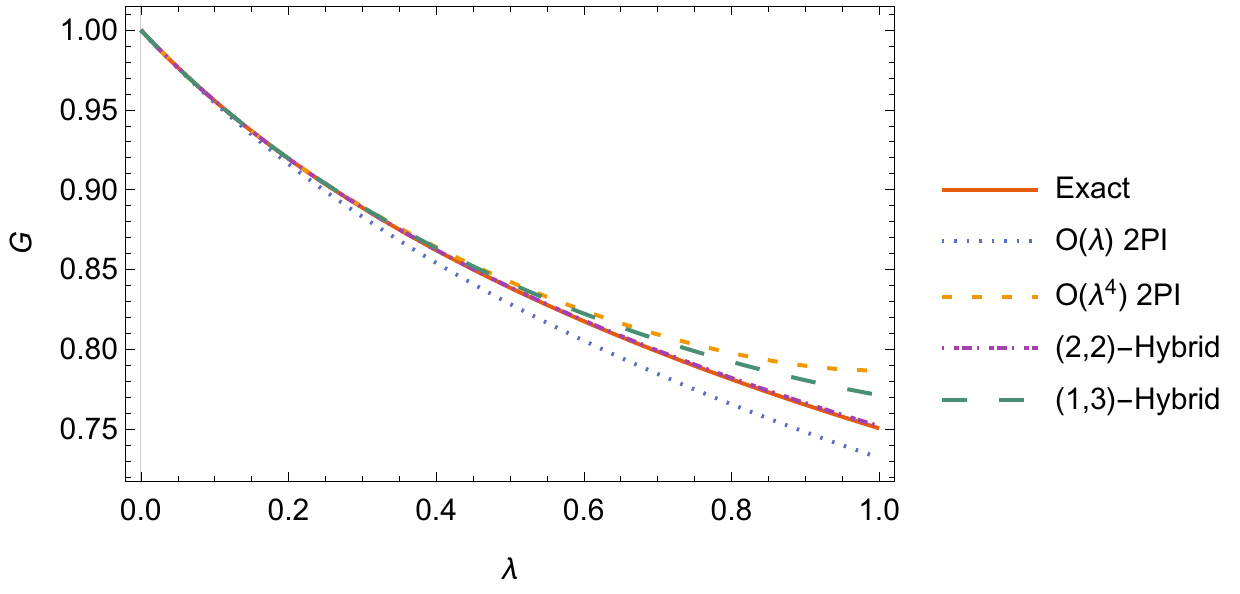}

\protect\caption{\label{fig:hybrid-solution}Comparison of exact $\bar{G}$, first
and fourth order 2PI and$\left(2,2\right)$- and $\left(1,3\right)$-2PI-Pad\'e
hybrid solutions at small coupling. The $\left(2,2\right)$-hybrid
solution lies almost on top of the exact solution.}
\end{figure}

\begin{figure}
\includegraphics[width=1\columnwidth]{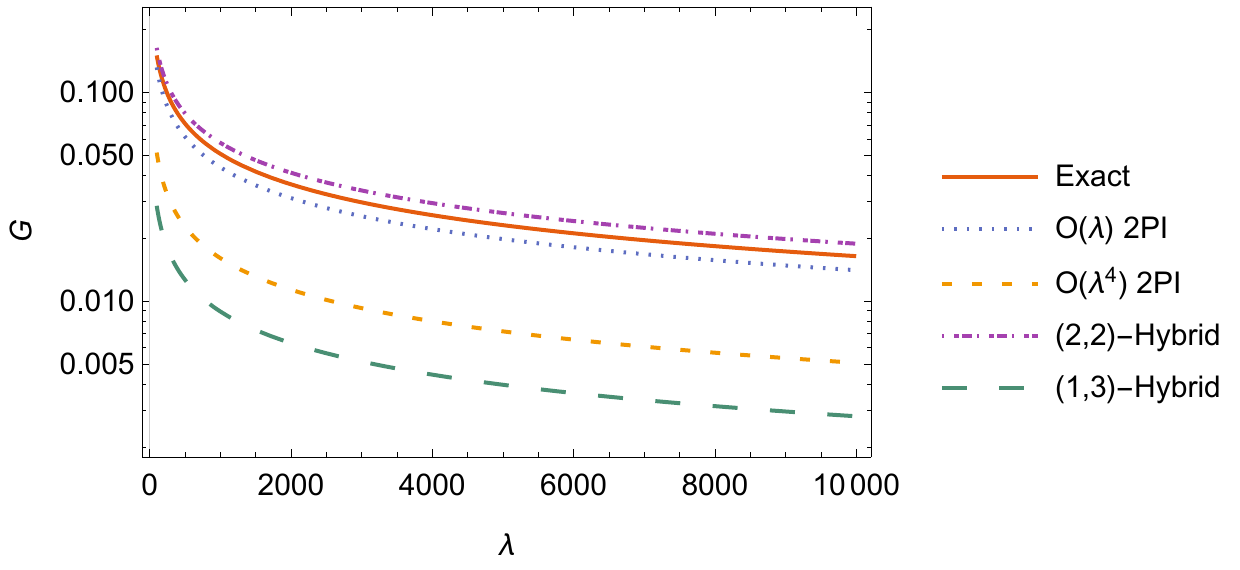}

\protect\caption{\label{fig:hybrid-solution-large-lambda}Comparison of exact $\bar{G}$,
first and fourth order 2PI and$\left(2,2\right)$- and $\left(1,3\right)$-2PI-Pad\'e
hybrid solutions at large coupling.}
\end{figure}

To take a specific example, consider
$\gamma_{2\text{PI}}$ up to $\mathcal{O}\left(\lambda^{4}\right)$
and the matching $\left(2,2\right)$- and $\left(1,3\right)$-Pad\'e
approximants:
\begin{align}
\gamma_{2\text{PI}} & =\frac{G^{2}\lambda}{8}-\frac{G^{4}\lambda^{2}}{48}+\frac{G^{6}\lambda^{3}}{48}-\frac{5G^{8}\lambda^{4}}{128}+\cdots\\
 & \approx\frac{\frac{G^{2}\lambda}{8}+\frac{113G^{4}\lambda^{2}}{480}}{1+\frac{41G^{2}\lambda}{20}+\frac{7G^{4}\lambda^{2}}{40}}\\
 & \approx\frac{G^{2}\lambda}{8\left(1+\frac{G^{2}\lambda}{6}-\frac{5G^{4}\lambda^{2}}{36}+\frac{113G^{6}\lambda^{3}}{432}\right)}.
\end{align}
The solution resulting from any of these versions of $\gamma_{2\text{PI}}$
cannot be written in closed form, however numerical solutions are
shown in Figure \ref{fig:hybrid-solution} for moderately small $\lambda$
(for $\lambda$ outside of this range, different roots become relevant).
We see that the hybrid solutions are more accurate than the fourth
order 2PI solution, although the level of improvement depends strongly
on the type of Pad\'e approximant employed. The trade-off is a loss
of accuracy at large $\lambda$, as shown in Figure \ref{fig:hybrid-solution-large-lambda}
for the best roots we can find. Each approximation decays with the
correct leading $\lambda^{-1/2}$ power, however they differ by $\mathcal{O}\left(1\right)$
factors which are not negligible. Remarkably, the best approximation
of those shown is the $\mathcal{O}\left(\lambda\right)$ 2PI approximation.
For the greater computational expense the $\left(2,2\right)$-hybrid
approximation is not noticeably an improvement at large coupling.
The apparent clustering of the $\mathcal{O}\left(\lambda^{4}\right)$
2PI and $\left(1,3\right)$-hybrid approximations away from the exact
value as $\lambda\to\infty$ probably stems from the minus sign of
the leading term in $\partial_{G}\gamma_{2\text{PI}}$ in these approximations,
suggesting that one should not use approximations with this property.

Now we try Pad\'e summing the equation of motion:
\begin{equation}
G^{-1}=m^{2}+2\frac{G^{-1}\sum_{n=1}^{N}A_{n}^{\left(\partial\gamma\right)}\left(\lambda G^{2}\right)^{n}}{\sum_{n=0}^{M}B_{n}^{\left(\partial\gamma\right)}\left(\lambda G^{2}\right)^{n}},\label{eq:hybrid-attempt-2}
\end{equation}
where the explicit factor of $G^{-1}$ on the right hand side simply
ensures the self-energy is an odd function of $G$ as it must be.
This equation of motion is of degree $\max\left(2N,2M+1\right)$,
which for typical choices of $N$ and $M$ results in a rough halving
of the number of spurious solutions. To obtain an analytic result
we consider the first non-trivial approximant, i.e. $N=M=1$. The
resulting equation of motion is
\begin{equation}
G^{-1}=m^{2}+\frac{1}{2}\frac{\lambda G}{1+\frac{1}{3}\lambda G^{2}},
\end{equation}
which agrees with the usual 2PI equation of motion up to terms of
order $\mathcal{O}\left(\lambda^{3}\right)$, however it only has
two spurious solutions instead of three. The physical solution obtained
by matching at small $\lambda$ is
\begin{equation}
G_{\text{hy}}=\frac{1}{6m^{2}}\left(-1+\frac{36m^{4}-\lambda}{X^{1/3}}-\frac{X^{1/3}}{\lambda}\right),\label{eq:hybrid-soln}
\end{equation}
where
\[
X=\lambda^{3}-378\lambda^{2}m^{4}+18\lambda m^{2}\sqrt{\lambda\left(-2\lambda^{2}+144m^{8}+429\lambda m^{4}\right)}.
\]

Indeed, this matches perturbation theory up to $\mathcal{O}\left(\lambda^{3}\right)$
terms. However, the large $\lambda$ behaviour in this approximation
is pathological:
\begin{equation}
G_{\text{hy}}\to-\frac{1}{2m^{2}}+\frac{18m^{2}}{\lambda}+\mathcal{O}\left(\lambda^{-3/2}\right),
\end{equation}
as $\lambda\to\infty$. This reflects the existence of an unphysical
branch cut on the positive $\lambda$ axis starting at $\lambda/m^{4}=\frac{3}{4}\left(143+19\sqrt{57}\right)\approx214$,
which $G_{\text{hy}}$ has in addition to the expected cuts on the
negative axis. The existence of this cut also renders the derivation
of the spectral function invalid, meaning that $G_{\text{hy}}$ cannot
be written in the form \eqref{eq:spectral-function-def}. It is not
clear at this stage whether other forms of Pad\'e summed equations of
motion lack these pathologies. We leave further investigation of hybrid
approximation schemes to future work.

\section{Discussion\label{sec:Discussion}}

Two-particle irreducible effective actions are the subject of a rich
literature and have found applications in diverse areas from early
universe cosmology to nano-electronics (e.g. \citep{Bergesa,Calzetta2008}).
The virtues of approximation schemes built on 2PI effective actions
are often explained in terms of a resummation of an infinite series
of perturbative Feynman diagrams which are encapsulated in the non-perturbative
Green function $G$, from which the 2PI diagram series is built. However,
the 2PI effective action is not a resummation scheme: it is a self-consistent
variational principle. The definition of the 2PI effective action
in terms of the Legendre transform is crucial for the self-consistency
of the scheme. The immediate practical consequence is that any modification
of the 2PI effective action which does not derive from a consistent
modified variational principle is very likely to be inconsistent.
So for instance, the consistency of recent attempts to improve the
symmetry properties of 2PI effective actions \citep{Pilaftsis2013}
is guaranteed by the existence of a suitable constrained variational
principle, however, ad hoc attempts to modify the equations of motion
to satisfy symmetries will fail.

In this work we have pointed out the distinction between resummation
and self-consistent approximations using an exactly solvable zero
dimensional ``field'' theory. The perturbation theory has zero radius
of convergence due to a branch cut on the negative coupling ($\lambda$)
axis, a fact which is invisible to perturbation theory. The theory
is Borel summable, with the Borel sum giving the exact answer. However,
Borel summation alone is not usually this helpful in practice. Pad\'e approximants
well describe the Green function at weak coupling, though not at strong
coupling. However, the combination of Borel and Pad\'e approximation
yields an effective global approximation scheme for the Green function.

The two point Green function of this theory admits a nice
integral representation in terms of the spectral function (i.e. discontinuity
of the branch cut) which allows us to see that the Pad\'e approximants
improve perturbation theory by allowing the spectral function to be
approximated as a sum of delta functions and the Borel-Pad\'e
method gives a continuous, albeit erratic and inaccurate,
approximation to the spectral function.

The 2PI approximation scheme surpasses perturbation theory,
Pad\'e and Borel-Pad\'e approximants already at the leading non-trivial truncation.
Like the Borel-Pad\'e method, 2PI approximations
can develop branch points and represent the spectral function by a
continuous distribution. However, the 2PI approximation is
quantitatively superior at the leading truncation.

We speculate that this is because the Borel-Pad\'e method
is a widely applicable general ``black box'' method, however the self-consistent
2PI equations of motion are derived within a \emph{particular} field theory
of interest. This gives the 2PI method ``insider information,'' from
which it should be able to construct a better approximation.
This comes at the cost of spurious solutions which must be eliminated and
the added difficulty of finding the 2PI effective action in the first
place.

Finally we introduced (for the first time, to our knowledge)
a hybrid 2PI-Pad\'e scheme, using Pad\'e approximants to partially resum
the 2PI diagram series. The quality of the result depends strongly
on the type of Pad\'e approximant used, with the best result in our
case for the diagonal approximant. This hybrid approximation performs
considerably better than the comparable 2PI approximation at weak
coupling, though not noticeably better at strong coupling.

During proofing we became aware of a work by Kleinert \citep{Kleinert1998} in some ways similar to our own.
In \citep{Kleinert1998}, Kleinert generalizes and reformulates the Feynman-Kleinert \emph{variational perturbation theory} and compares it to the 2PI effective action (called by the non-standard name ``bilocal Legendre transform'') for the toy model discussed here. He finds that the variational perturbation theory out-performs the 2PI method and notes especially the failure of the 2PI series to converge uniformly for all couplings (this is not a surprise in light of the fact that the 2PI series is asymptotic).
Our work differs in scope from his in two major ways. First, we compare the 2PI method to traditional resummation methods and not to other variational methods. Second, Kleinert focuses on the value of the effective action itself evaluated at its extremum, while we focus on the correlation function. In particular, we have focused on the analytic structure of the correlation function in the complex coupling plane, and introduced the spectral representation to aid this discussion. A study of the behaviour of these quantities in variational perturbation theory is an intriguing prospect for future work. (An intermediate result of Kleinert's work is directly applicable to our discussion around \eqref{eq:gamma2pi-series}: equation 110 of \citep{Kleinert1998} gives a non-linear recurrence relation for the coefficients of the 2PI self-energy.)

There are several other natural directions for extension of this work. One
would be the calculation and comparison of higher orders in the 2PI
and (Borel-)Pad\'e approximations, although it is doubtful what new insights
might come from this. It would be straightforward to extend this work
to consider 4PI effective actions, which depend on a self-consistent
vertex function $V^{\left(4\right)}$ in addition to $G$.\footnote{By replacing $\lambda\to\lambda+L$ in \eqref{eq:partition-function}
and performing the Legendre transform with respect to both $L$ and
$K$.} However, the extension to higher order $n$PI effective actions requires
the introduction of new terms in the exponent of \eqref{eq:partition-function}
which makes the problem no longer exactly solvable.
A natural direction to pursue is Borel-Pad\'e summation of the 2PI generating
functional itself.  This opens the possibility of eliminating spurious solutions
to the 2PI equations of motion and enhancing the sensitivity of the 2PI
method to the exponentially small regions of the spectral function, hence
eliminating unphysical cusps.
We are planning an investigation of these themes for the quantum anharmonic oscillator
as a stepping stone to more physically interesting field theories.


\end{document}